%% file: costa.tex
\documentclass[]{aastex63}
\usepackage{natbib}
\usepackage{multirow}
\usepackage[caption=false]{subfig}
\include{defs}
\usepackage{graphicx}
\usepackage{amsmath}
\usepackage[normalem]{ulem}
\usepackage{ mathrsfs }
\usepackage{hyperref}

\usepackage{threeparttable}
\usepackage{booktabs}

\newcolumntype{P}[1]{>{\centering\arraybackslash}p{#1}}
\newcommand{\mytablefill}{...}

\begin{document}

\title{Towards a More Complex Understanding of Natal Super Star Clusters with Multiwavelength Observations}

\author[0000-0002-7408-7589]{Allison H. Costa}
\affiliation{Department of Astronomy, University of Virginia, Charlottesville, Virginia} 

\author{Kelsey E. Johnson}
\affiliation{Department of Astronomy, University of Virginia, Charlottesville, Virginia}

\author{Remy Indebetouw}
\affiliation{Department of Astronomy, University of Virginia, Charlottesville, Virginia}\affiliation{National Radio Astronomy Observatory, Charlottesville, Virginia}

\author{Molly K. Finn}
\affiliation{Department of Astronomy, University of Virginia, Charlottesville, Virginia}

\author{Crystal L. Brogan}
\affiliation{National Radio Astronomy Observatory, Charlottesville, Virginia}

\author{Amy Reines}
\affiliation{Montana State University, Bozeman, Montana}

\begin{abstract}
Henize 2-10 (He 2-10) is a nearby (D = 9 Mpc) starbursting blue compact dwarf galaxy that boasts a high star formation rate and a low luminosity AGN \citep{Reines:2011, Reines:2016}. He 2-10 is also one of the first galaxies in which embedded superstar clusters (SSCs) were discovered. SSCs are massive, compact star clusters that will impact their host galaxies dramatically when their massive stars evolve. Here, we discuss radio, submillimeter, and infrared observations of He 2-10 from 1.87 $\mu$m to 6 cm in high angular resolution ($\sim$0.3\arcsec), which allows us to disentangle individual clusters from aggregate complexes as identified at lower resolution. These results indicate the importance of spatial resolution to characterize SSCs, as low resolution studies of SSCs average over aggregate complexes that may host SSCs at different stages of evolution. We explore the thermal, non-thermal, and dust emission associated with the clusters along with dense molecular tracers to construct a holistic review of the natal SSCs that have yet to dramatically disrupt their parent molecular clouds. We assess the production rate of ionizing photons, extinction, total mass, and the star formation efficiency associated with the clusters. Notably, we find that the star formation efficiency for the some of the natal clusters is high ($>$ 70$\%$), which suggests that these clusters could remain bound even after the gas is dispersed from the system from stellar feedback mechanisms. If they remain bound, these SSCs could survive to become objects indistinguishable from globular clusters. 

\end{abstract}

\keywords{}

\section{Introduction}

Rich clusters of stars significantly impact their host galaxies, as the massive stars can undergo nearly simultaneous supernovae, driving outflows into the interstellar medium and potentially into the intergalactic medium \citep{Johnson:2000, Heckman:2001, Martin:2002,Gilbert:2007}. Super star clusters (SSCs) are an example of some of the most extreme types of stellar systems with stellar densities that can be orders of magnitudes higher than a typical open cluster. We adopt as description of SSCs that they are massive ($\gtrsim$ 10$^{4.5}$ \Msun) compact ($r$ $\leq$ 3 pc) star clusters with stellar densities often exceeding 10$^4$ stars pc$^{-3}$ in their cores (\citealt{Ryon:2017} and references therein; \citealt{Johnson:2018}). 

Merging and starbursting galaxies in the current universe may host similar environments to those that formed globular clusters (GCs), e.g., gas-rich, high pressure disks, and turbulent \citep{Portegies:2010,Keller:2020}, and SSCs are found in such environments. Whether SSCs are consistent with being progenitors of GCs has been an ongoing line of study (e.g., \citealt{Holtzman:1992,Whitmore:1995,Whitmore:2000,Harris:2003,Portegies:2010,Longmore:2014,Horta:2021}).  If SSCs are local analogs to GC precursors, then nearby SSCs provide an opportunity to study the physical conditions necessary for GCs to form. The infant mortality rate of GCs is likely to be quite high (possibly as high as 99$\%$; \citealt{Fall:2001}), which suggests that this mode formation may have been critical, and perhaps dominant, in the time of galaxy assembly in order to account for the population of GCs we observe today.  Recent simulations (e.g., \citealt{Keller:2020,Horta:2021}) suggest that GC formation is not primarily due to an exotic formation scenario (e.g., dark matter mini halos) but instead it is the extreme end of a continuous distribution. Observations of SSCs in merging or starbursting galaxies can provide insight to the formation process, the conditions of the local interstellar medium, and perhaps what characteristics are important to enable a cluster to remain bound such that it will persist for billions of years. While much is known about adolescent SSCs with ages between $\sim$3 -- 10 Myr, these clusters have already emerged from their birth material to be clearly visible at optical wavelengths, thus erasing their birth conditions as the clusters themselves have already undergone significant evolution. We need to probe SSCs which have yet to emerge from their parent molecular environments to fully characterize this mode of star formation.

The starbursting dwarf galaxy Henzie 2-10 (He 2-10; R.A. = 08$^h$ 36$^m$ 15.120$^s$, decl = --26\ddeg{} 24\arcmin{} 34.157\arcsec) is one of the nearest galaxies (D = 9 Mpc; \citealt{Johansson:1987}) known to host a multitude of SSCs at varying stages of evolution. From Hubble Space Telescope (HST) data, \citet{Johnson:2000} determined masses for the optically visible adolescent SSCs in He 2-10. They found that SSCs in the central region of He 2-10 (their region `A') have masses between 1.6 -- 2.6 $\times$ 10$^6$ \Msun{} while adolescent SSCs in a region $\sim$ 9\arcsec{} east have lower masses between 2.6 -- 6.6 $\times$ 10$^4$ \Msun{} (their region `B'). Within $\sim$ 5\arcsec{} of the nominal center of He 2-10, \citet{Kobulnicky:1999} identified radio sources at 6 cm that are consistent with being Ultra Dense (UD) \HII regions. UD \HII regions are dense (n$_e$ $\gtrsim$ 10$^4$ cm$^{-3}$) ionized structures of a few parsecs in size that host a massive stellar cluster \citep{Kobulnicky:1999,Vacca:2002}. They are extragalactic scaled up versions of galactic Ultra Compact (UC) \HII regions, which are typically excited by a single massive star and are $\lesssim$  0.1 pc in size  \citep{Wood:1989}. \citet{Kobulnicky:1999} suggested further that these UD \HII regions host the young precursors to optically visible SSCs: natal SSCs.

\begin{figure}[htb!]
\centering
\includegraphics[width=0.7\textwidth]{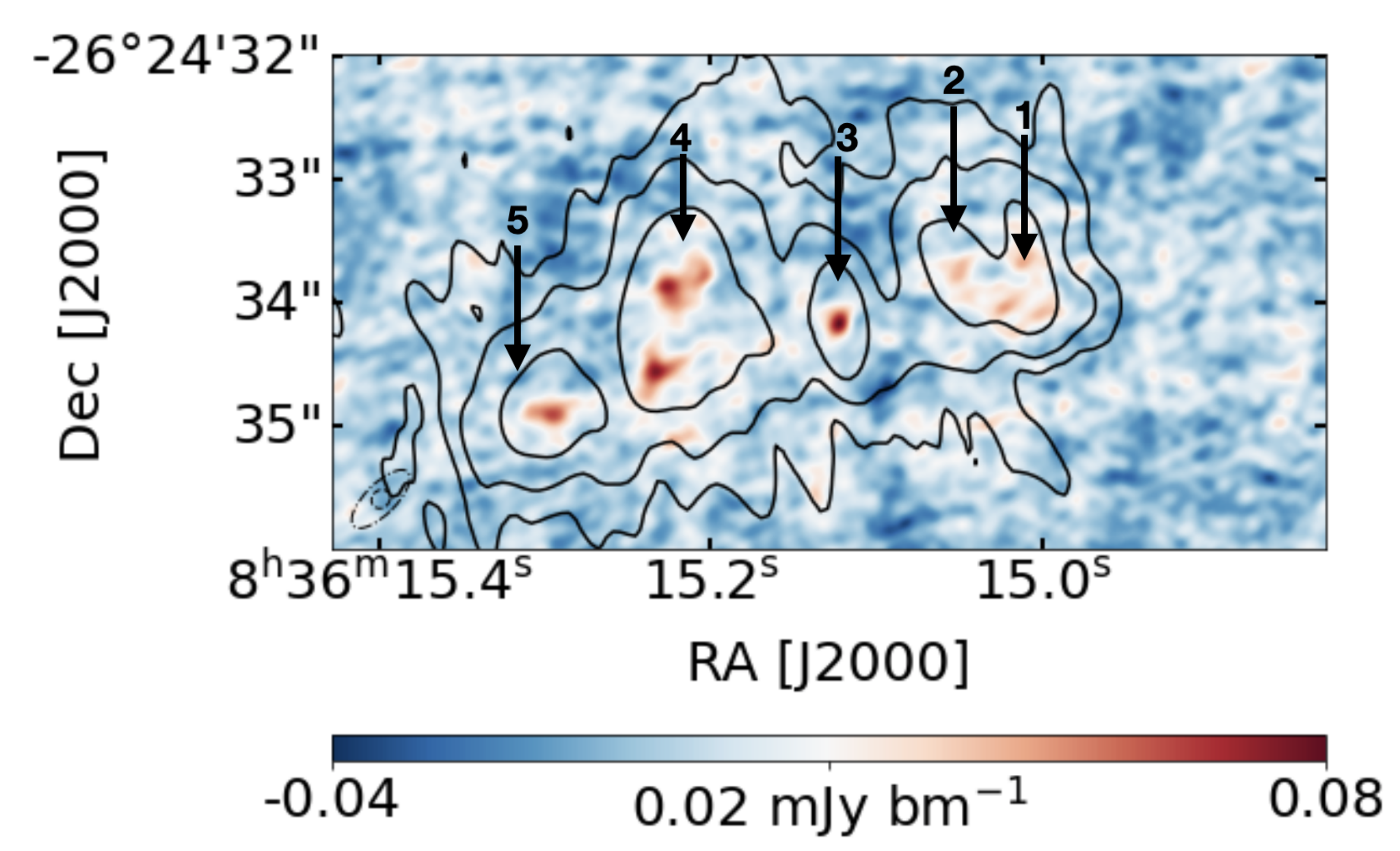}
\caption{33 GHz continuum map of Henize 2-10. The contours are the 5 GHz map from VLA/AJ314 to show radio knots as originally classified by \citet{Kobulnicky:1999}. The 5 GHz contours are at 5, 10, and 22$\sigma$. The beams are shown in the lower left.}
\label{fig:radioknots}
\end{figure}

\citet{Kobulnicky:1999} labeled these radio sources Knots 1, 2, 3, 4, and 5 (Figure \ref{fig:radioknots}), and since their identification, the radio knots have inspired multiwavelength studies to assess the properties of the natal SSCs and their environments. With the historical Very Large Array, \citet{Johnson:2003} modeled the spectral energy distributions (SEDs) of the knots between 5 and 43 GHz; they estimated stellar masses of (1.2, 1.6, 4.3, and 1.3) $\times$ 10$^{5}$ \Msun{} for Knots 1, 2, 4, and 5, respectively, and modeled the physical properties of the UD \HII regions. From infrared (IR) observations with the Very Large Telescope (VLT) at 2.2, 3.8, and 4.8 $\mu$m as well as HST data, \citet{Cabanac:2005} confirmed the classification of Knots 1, 2, and 5 as UD \HII regions and reported that the clusters are young ($\lesssim$ 6 Myr). \citet{Cabanac:2005} identified a number of optical sources spatially coincident with Knot 4, labeled in their Figure 2a as L4a -- L4d, which they interpreted as Knot 4 being a mix of normal \HII regions and supernova remnants (SNRs) instead of an UD \HII region. \citet{Reines:2012} also reported a SNR in Knot 4 that is spatially coincident with the most optically-luminous SSC in their Very Long Baseline Interferometry (VLBI) observations.

\citet{Johnson:2003} found that Knot 3 had a spectral index consistent with having significant non-thermal emission, and in more recent work, \citet{Reines:2011}, \citet{Reines:2012}, \citet{Reines:2016}, and \citet{Riffel:2020} identified the low luminosity central black hole of He 2-10 hosted in Knot 3 from VLA, VLBI, and \textit{Chandra} observations. \citet{Hebbar:2019} suggest Knot 3 may be a supernova remnant based on their reanalysis of the X-ray spectrum first presented by \citet{Reines:2016}. They argue that a hot plasma model fits the X-ray spectrum better than a power-law model, with the former and latter typically used for supernova remnants and luminous AGNs, respectively. However, \citet{Reines:2016} note that the soft X-ray spectrum is similar to other low-luminosity AGNs radiating at low Eddington ratios \citep{Constantin:2009,Baganoff:2003}.  The X-ray spectrum also shows evidence for variability on hour-long timescales, which is incompatible with a supernova remnant \citep{Reines:2016}.

He 2-10 also hosts dense molecular precursor clouds, which may have the potential to form SSCs. \citet{Johnson:2018} mapped CO with the Submillimeter Array (SMA) and dense molecular tracers such as HCN, HNC, CCH, and HCO$^+$ in He 2-10 with the \ALMA. They discuss the molecular morphology of these dense clouds and how the molecular lines are correlated with the evolutionary state of the natal clusters. Near the optical SSCs and the AGN, there is a a valley in CO and dense gas tracers \citep{Johnson:2018, Beck:2018}; however, the embedded natal SSCs appear to coincide with peaks in the denser molecular tracers \citep{Johnson:2018}. \citet{Imara:2019} also presented a study of CO to characterize differences in stellar nurseries between He 2-10 and the Milky Way. They reported that the most massive Giant Molecular Clouds He 2-10 are gravitationally bound, which suggests that He 2-10 can continue to form SSCs. He 2-10 hosts natal and adolescent SSCs and has the potential to continue to form them, making it an ideal laboratory to understand this mode of star formation.

In this work, we discuss the results of radio, mm, submm, and IR observations of the radio knots in He 2-10 with high resolution data that allow us to disentangle individual clusters from aggregate complexes as identified at lower resolution (e.g., \citealt{Kobulnicky:1999}). We present archival Karl G. Jansky Very Large Array\footnote{\vlatext} (VLA) and ALMA data to construct SEDs of the radio knots to characterize the environments that host the natal SSCs. In Section \ref{sec:kaobs}, we present new VLA 33 GHz observations of He 2-10 from VLA program 15B-197.  In addition, there is a wealth of continuum data not yet analyzed in the VLA and ALMA archives, though some of the data have been published for other purposes (e.g., spectral lines). In Section \ref{sec:vlaold}, we discuss pre-upgrade VLA 5 and 8 GHz observations and more recent 15 and 22 GHz archival observations. In Section \ref{sec:almaobs}, we present an overview of the archival observations of He 2-10 with ALMA, and in Section \ref{sec:nicmosobs}, we discuss archival HST/NICMOS observations at 1.8756 $\mu$m. Section \ref{sec:photom} presents the source finding and photometry of the radio knots. In Section \ref{sec:SED}, we discuss the SED model and results. We present estimates of the production rate of the ionizing photons and the stellar masses of the sources in Section \ref{sec:qly} and in Section \ref{sec:extinction}, the extinction estimates. Section \ref{sec:dustmass} presents our analysis of the dust mass associated with the knots, and Section \ref{sec:SFE} discusses the star formation efficiency of the knots.  Finally in Section \ref{sec:dis}, we provide an overview of our results and characteristics of the radio knots of He 2-10.

\section{Observations and Data Reduction\label{sec:obs}}

A range of new and archival long wavelength data are available from observations spanning the last two decades. These data have not yet been systematically evaluated and analyzed in conjunction. Here, we summarize the archival data and recent observations of He 2-10 at radio, mm, and submm wavelengths.

\subsection{New VLA Observations\label{sec:kaobs}}

With the VLA, we observed He 2-10 at Ka-band (28$\;$GHz -- 36$\;$GHz) in A configuration between August and September of 2015. Table \ref{tab:obs} gives the details of the observations. We used the source J1331+3030 (3C286) as the flux density calibrator, J1229+0203 (3C273) as the bandpass calibrator, and J0836-2016 as the complex gain calibrator. Due to the high frequency of the observations and array configuration, scans of the complex gain calibrator were interleaved with scans of the program source in intervals of $\sim$2 minutes and pointing scans were completed every hour. With the 3-bit correlator configuration, we obtained a total bandwidth of $\sim$8\ghz{} across Ka-band.

\begin{table}[htb!]
\centering
\begin{threeparttable}
\centering
\caption{Log of Observations\label{tab:obs}}
\begin{tabular}{p{8cm}p{6cm}}
\toprule
VLA Project Code & 15B-197 \\
Date of Observations & 2015 Aug 31; 2015 Sept 1, 8, 19, $\&$ 24 \\
Number of Scheduling Blocks & 4\\
Duration of Scheduling Blocks (hr) & 1.5 \\
Frequencies Observed (GHz)\tnote{a} & 30; 32; 34; 36 \\
Number of Frequency Channels per IF & 2048 \\
Channel Width (MHz) & 1 \\
VLA configuration & A \\
Total Integration Time on Source (hr) & 2.4 \\
\bottomrule
\end{tabular}
\begin{tablenotes}
\item[a] \small{These were 3-bit observations that had 2.048$\;$GHz wide intermediate frequency bands (IFs) centered on the frequencies listed. Each IF was composed of sixteen 128$\;$MHz wide subbands.}
\end{tablenotes}
\end{threeparttable}
\end{table}

The data were reduced and imaged using the NRAO Common Astronomy Software Applications (CASA) version 5.4.2 (\url{casa.nrao.edu}; \citealt{McMullin:2007}). We checked the flux calibration by using 3C286 as the flux and bandpass calibrator and calibrating the 3C273 as a science target; the reported flux density of 3C273 was consistent with the VLA Calibrator Manual\footnote{\url{https://science.nrao.edu/facilities/vla/observing/callist}}, and the two methods of calibration produced similar results within the uncertainties. Four 1.5 hr observations were completed in A array yielding $\sim$2.4 hr integrated time on source\footnote{Six additional observations were completed while the VLA configuration was transitioning to D array in September 2015. The resulting u-v coverage in these latter observations proved to be un-imageable because the u-v coverage was so irregular. Thus, the data in these 6 scheduling blocks were discarded.}. \textit{Natural} weighting was employed in the CASA task {\sc tclean} to recover the best signal/noise ratio possible at the expense of resolution. He 2-10 is a low declination source ($\sim$ --26\ddeg), so we down-weighted the E-W arms and used a taper of 800k$\lambda$ $\times$ 4000k$\lambda$ $\times$ 90\ddeg, resulting in a synthesized beam of 0.162\arcsec{} $\times$ 0.140\arcsec. While we do achieve the theoretical rms for the 2.4 hr integration time, there are still significant artifacts in the map despite these efforts. High frequency observing is sensitive to the weather conditions, and given the low declination of the source, we believe the data suffer from significant decorrelation. Thus, the flux densities reported here are treated as a lower limit thorough this work.

\subsection{VLA Archival Observations\label{sec:vlaold}}

At longer radio wavelengths, we selected the newest observations of the appropriate resolution from the VLA archive. The most recent 4 GHz (C-band) and 8 GHz (X-band) observations available for He 2-10 are in program AJ314, which are from 2004 and before the VLA was upgraded. These observations included the Pie Town up-link. These data were initially published in \citet{Reines:2011} with focus on properties of the AGN. The flux density calibrator was 3C286, and the complex gain calibrator was  J0836-2016. The data were re-imaged using CASA 5.4.2. Table \ref{tab:alldataprop} lists the final beam, the peak intensity, and the rms of the final images for this program and the ones that follow below. VLA program 16B-067 contains 14 GHz (Ku-band) and 22 GHz (K-band) continuum data, which we reprocessed  and imaged using the VLA pipeline in CASA 5.4.2. We applied a taper of 600k$\lambda$ $\times$ 3000k$\lambda$ $\times$ 90\ddeg{} during imaging in both cases. These data have not been published to our knowledge. For these data, 3C138 was the flux density and bandpass calibrator, and J0836-2016 was the complex gain calibrator.

\subsection{ALMA Archival Observations\label{sec:almaobs}}

Here, we detail the archival ALMA observations that we include in our analysis of He 2-10. The calibrated data were retrieved from the archive and restored using the delivered scripts from NRAO. Properties of the images are given in Table \ref{tab:alldataprop}.

The $^{12}$CO$\;$(1-0) and 113$\;$GHz continuum data were obtained for the related projects 2015.1.01569 and 2016.1.00027, with an originally proposed resolution of 0.6\arcsec.  Fortunately, the 2016.1.00027 data were taken while ALMA was between configurations, so some longer baselines were included in the array at the time of observation. Thus, we were able to re-image the data at higher angular resolution. 

Project 2015.1.01569 was observed twice with 39 antennas each time; on 2016 July 31, J1037-2934 (900 mJy) was used as the amplitude and bandpass calibrator and J0826-2230 (750 mJy) as the phase calibrator. On 2016-08-02, J1107-4449 (1.1 Jy), J1037-2934, and J0826-2230 were used as the amplitude, bandpass, and phase calibrators, respectively. The data were calibrated with the ALMA data pipeline Pipeline-Cycle3-R4-B, v.36660 \citep[\url{https://almascience.nrao.edu/processing/science-pipeline}][]{Davis:2020}, packaged with CASA 4.5.3. These data were initially published in \citet{Imara:2019}.

Project 2016.1.00027 was observed twice on 2017 July 13 with 42 antennas; the first time was using J0538-4405 (1.2 Jy) as the amplitude calibrator, J1037-2934 (1.3 Jy) as the bandpass calibrator, and J0846-2607 (330mJy) as the phase calibrator. The second execution used J1037-2934 as the amplitude and bandpass calibrator and J0846-2607 as the phase calibrator. The data were calibrated with Pipeline-Cycle4-R2-B v.39732 packaged with CASA 4.7.2 and imaged in CASA 5.6.1-8.

At 250 GHz, project 2012.1.00413 was observed 5 times between 2015 Sept 05 and 2015 Sept 29 with 30-35 antennas.  J1037-295 (650 mJy) or J0538-4405 (1.3 Jy) were used for the amplitude calibration, J0750+1231 (570 mJy), J0538-4405, or J1058+0133 (3.3 Jy) for the bandpass calibration, and J0747-3310 (500 mJy) or J0826-2230 (300-450 mJy) for the phase calibration. The data were calibrated with Pipeline-Cycle3-R1-B v.34044 in CASA 4.3.1 and imaged in CASA 5.6.1-8. These data were initially published in \citet{Johnson:2018}.

Finally at 340 GHz, project 2016.1.00492 was observed on 2016-11-15, using J0538-4405 (340 mJy) for the amplitude and bandpass calibration and J0846-2607 (210 mJy) for the phase calibration. The data were calibrated with Pipeline-Cycle4-R2-B v.38377 in CASA 4.7.0 and re-imaged with CASA 5.6.1-8. The data were initially published in \citet{Beck:2018} with a focus on the $^{12}$CO$\;$(3-2) data. 

\begin{table}
\centering
\setlength\tabcolsep{7pt}
\begin{threeparttable}
\centering
\caption{Log of Projects Used in this Work\label{tab:alldataprop}}
\begin{tabular}{lccccc}
\toprule
\multirow{2}{*}{Project} & $\nu$ & \multirow{2}{*}{Robust} & Beam    & Peak & rms \\
	& (GHz)	& 	 & (\arcsec) & (mJy bm$^{-1}$) & (mJy bm$^{-1}$) \\
\midrule
\multirow{2}{*}{VLA AJ314\tnote{a}}	  & \multirow{2}{*}{5}	& \multirow{2}{*}{0.0} & 0.620 $\times$ 0.248	& 1.06 & 0.015 \\
    		  &     &     & 0.300		& 0.95 & 0.018 \\
 \hline
\multirow{2}{*}{VLA AJ314}        & \multirow{2}{*}{8} & \multirow{2}{*}{0.5} & 0.366 $\times$ 0.150	& 0.57 & 0.012 \\
    		  &     &     & 0.300		& 0.65 & 0.012 \\
    		   \hline
\multirow{2}{*}{VLA 16B-067} & \multirow{2}{*}{15} &\multirow{2}{*}{0.5} & 0.322 $\times$ 0.210 & 0.35 & 0.017 \\ 
    		  &     &     & 0.300		&  0.39 & 0.024 \\
    		   \hline
\multirow{2}{*}{VLA 16B-067} & \multirow{2}{*}{22} &\multirow{2}{*}{0.5} & 0.192 $\times$ 0.152 & 0.31 & 0.023 \\ 
    		  &     &     & 0.300		& 0.48 & 0.023 \\
    		   \hline
\multirow{2}{*}{VLA 15B-197} & \multirow{2}{*}{33} &\multirow{2}{*}{2.0\tnote{b}} & 0.162 $\times$ 0.140 & 0.08 & 0.012\\ 
    		  &     &     & 0.300		& 0.16 & 0.019 \\
    		   \hline

\multirow{2}{*}{ALMA 2016.1.00027} & \multirow{2}{*}{113} & \multirow{2}{*}{0.0} & 0.381 $\times$ 0.206	& 0.30 & 0.017 \\
	     	  & 	&     & 0.300		& 0.29 & 0.019 \\
	     	   \hline
\multirow{2}{*}{ALMA 2012.1.00413} & \multirow{2}{*}{250} & \multirow{2}{*}{2.0} & 0.191 $\times$ 0.157	& 0.14 & 0.013 \\
	     	  & 	&     & 0.300		& 0.28 & 0.025 \\
	     	   \hline
\multirow{2}{*}{ALMA 2016.1.00492} & \multirow{2}{*}{340} & \multirow{2}{*}{0.5} & 0.306 $\times$ 0.265    	& 0.39 & 0.052 \\
	     	  & 	&     & 0.300		& 0.47 & 0.048 \\
\hline\hline
HST/NICMOS & $\lambda_{peak}$ & \multirow{2}{*}{\mytablefill} & FWHM & 0.42 & 0.014 \\
 F187N & 1.8756  $\mu$m & & 0.0191 $\mu$m & mJy pixel$^{-1}$ & mJy pixel$^{-1}$ \\
	     	   \bottomrule
\end{tabular}
\begin{tablenotes}
\item[a] The top row associated with each project code lists the image properties, and the following row gives the image properties of the convolved map.
\item[b] Robust of 2.0 is nearly equivalent to \textit{natural} weighting.
\end{tablenotes}
\end{threeparttable}
\end{table}

\subsection{HST NICMOS Archival Observations \label{sec:nicmosobs}}

The Near Infrared Camera and Multi-Object Spectrometer (NICMOS) data from our project 10894 were retrieved from the Hubble Legacy Archive. These data were initially published in \citet{Reines:2011}, which focused on properties of the AGN.  At He 2-10’s radial velocity of 873 \kms, the narrow-band filters F190N and F187N have transmission at Paschen $\alpha$ (\palpha) of 0.016 and 0.22 (close to the maximum for that filter), respectively. The images required a shift of 1.33\arcsec{} to match the radio data; we estimate a positional uncertainty of 0.1\arcsec. The images were converted to flux density units (Jansky pixel$^{-1}$). If the SEDs of the SSC are Rayleigh-jeans tails of hot star SEDs, the continuum in F187N will be 3$\%$ brighter than in F190N.  If the SED shape is different, for example with a hot dust contribution, that number will differ. In the subsequent analysis, we scaled the F187N filter down by the theoretical scale of 3$\%$ before subtracting it from the F190N filter.

\subsection{Source Extraction and Photometry\label{sec:photom}}

\begin{figure}[!htb]
\centering
\gridline{\fig{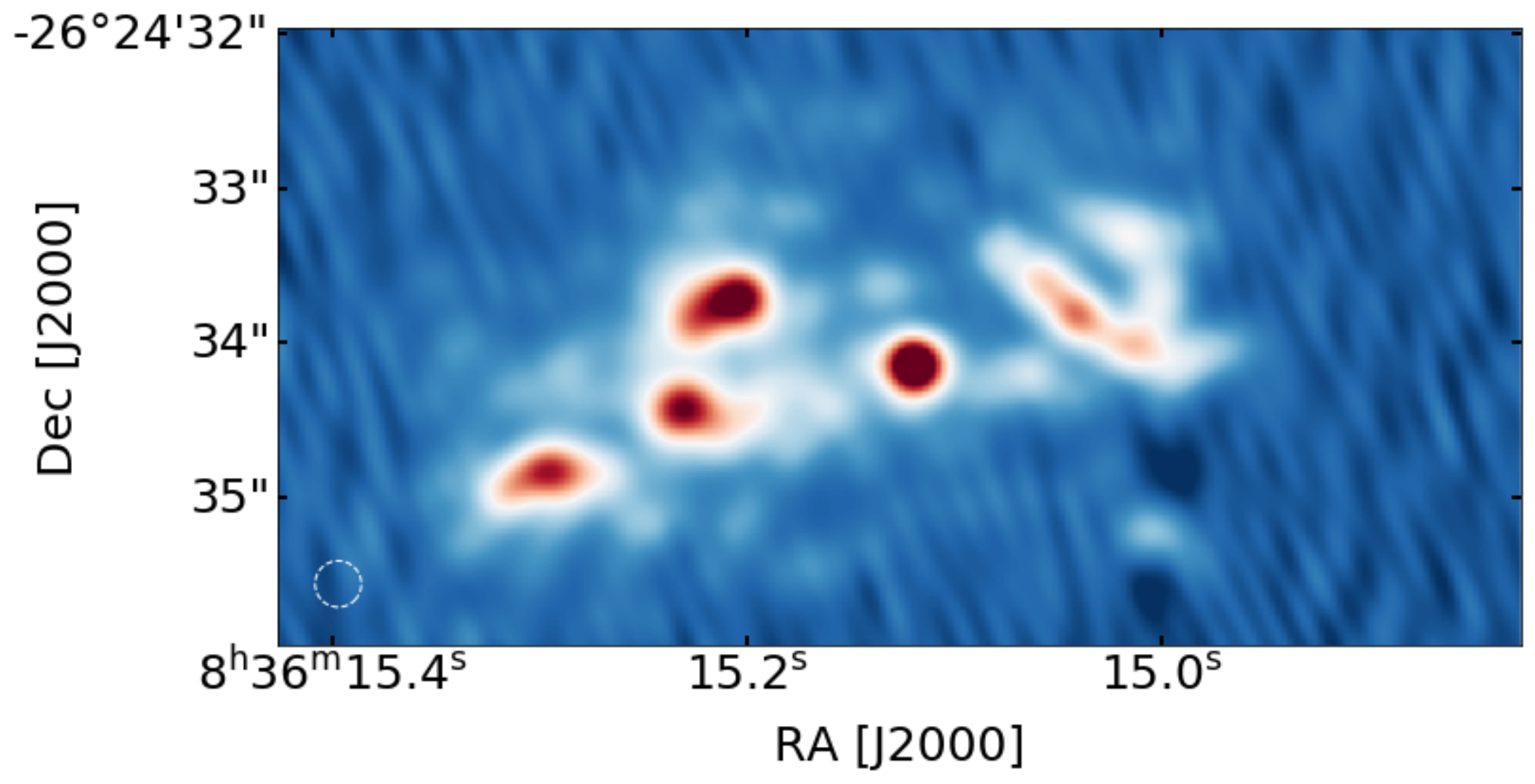}{0.49\textwidth}{(a) 5 GHz}
          \fig{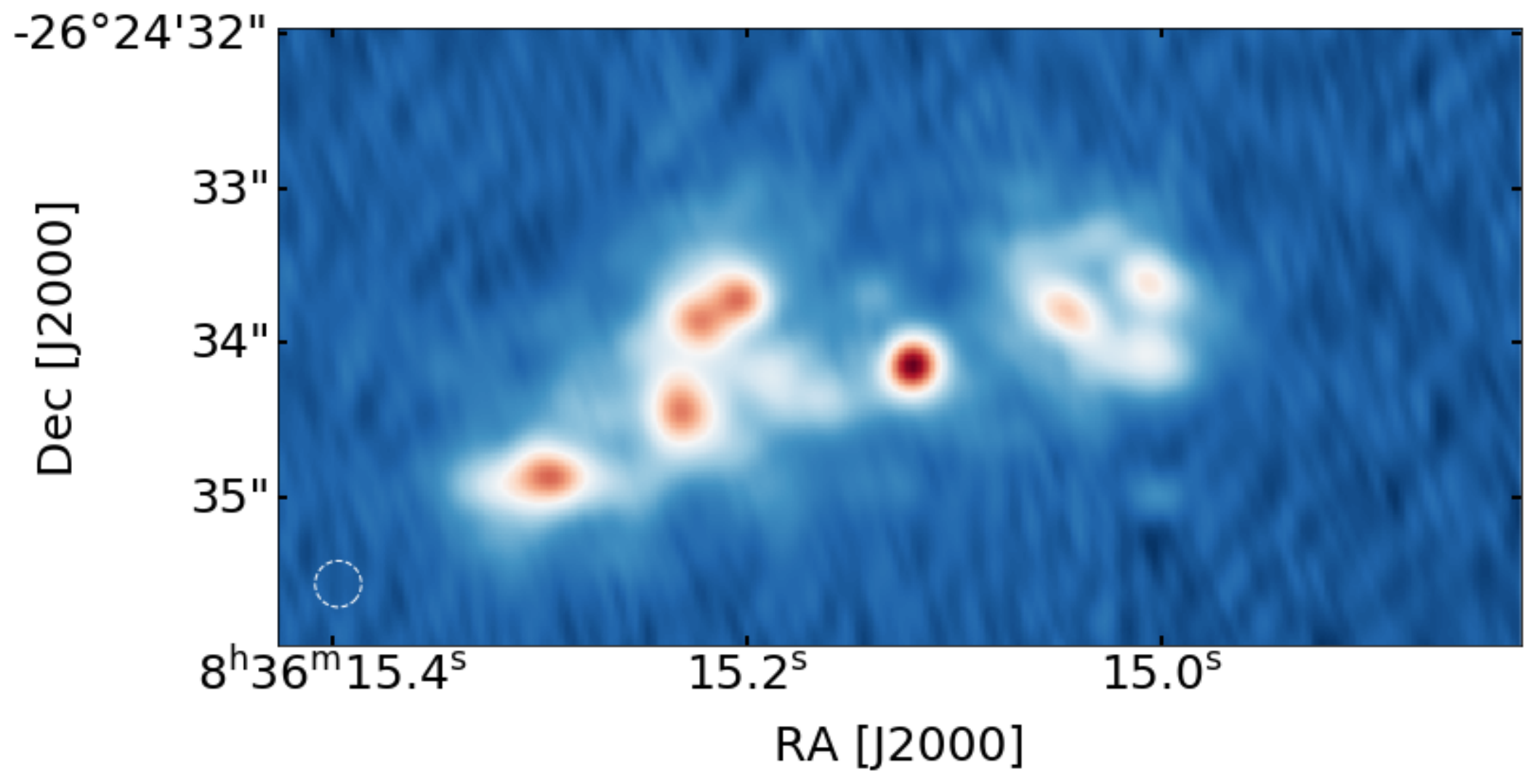}{0.49\textwidth}{(b) 8 GHz}}
          
\gridline{\fig{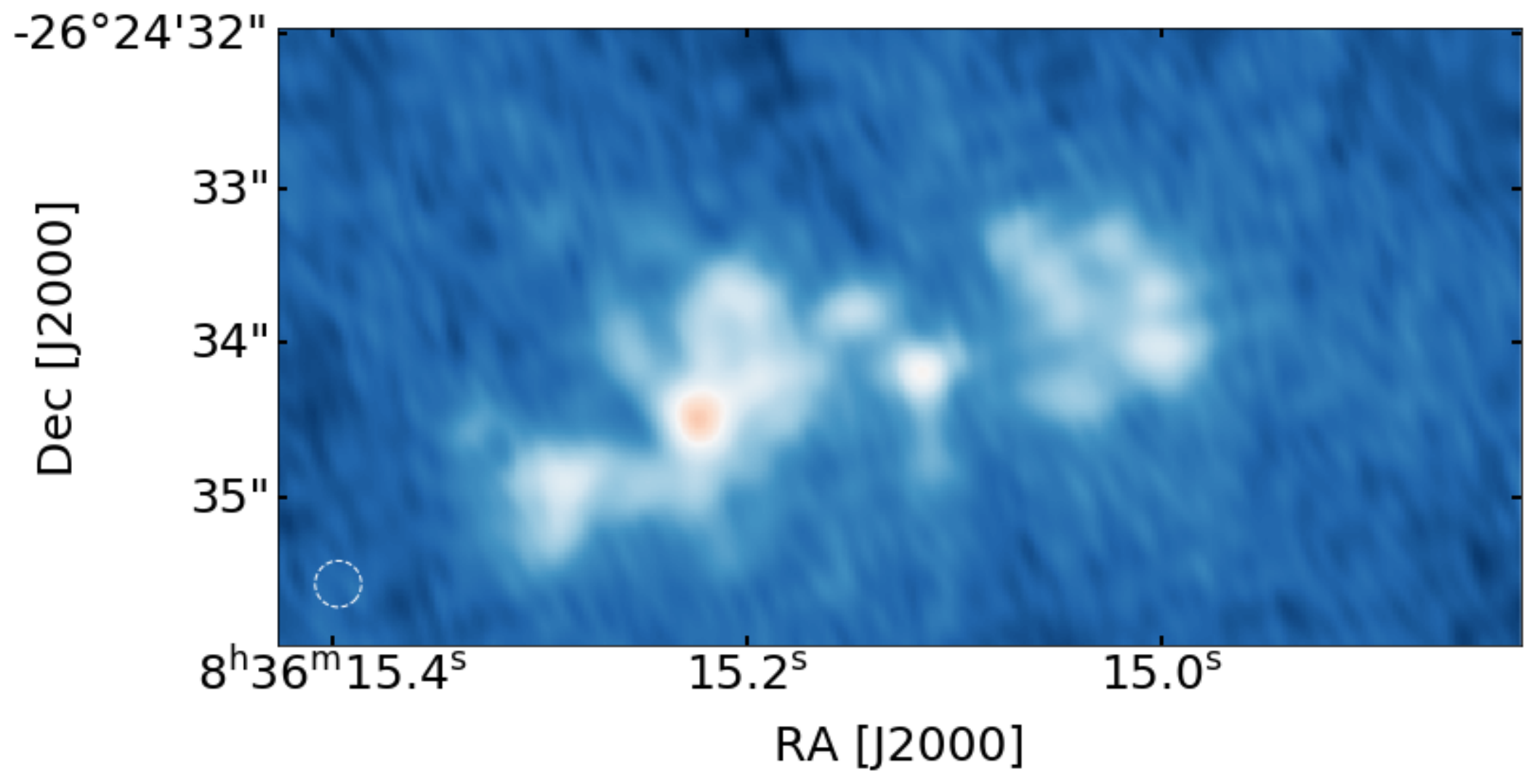}{0.49\textwidth}{(c) 15 GHz}
          \fig{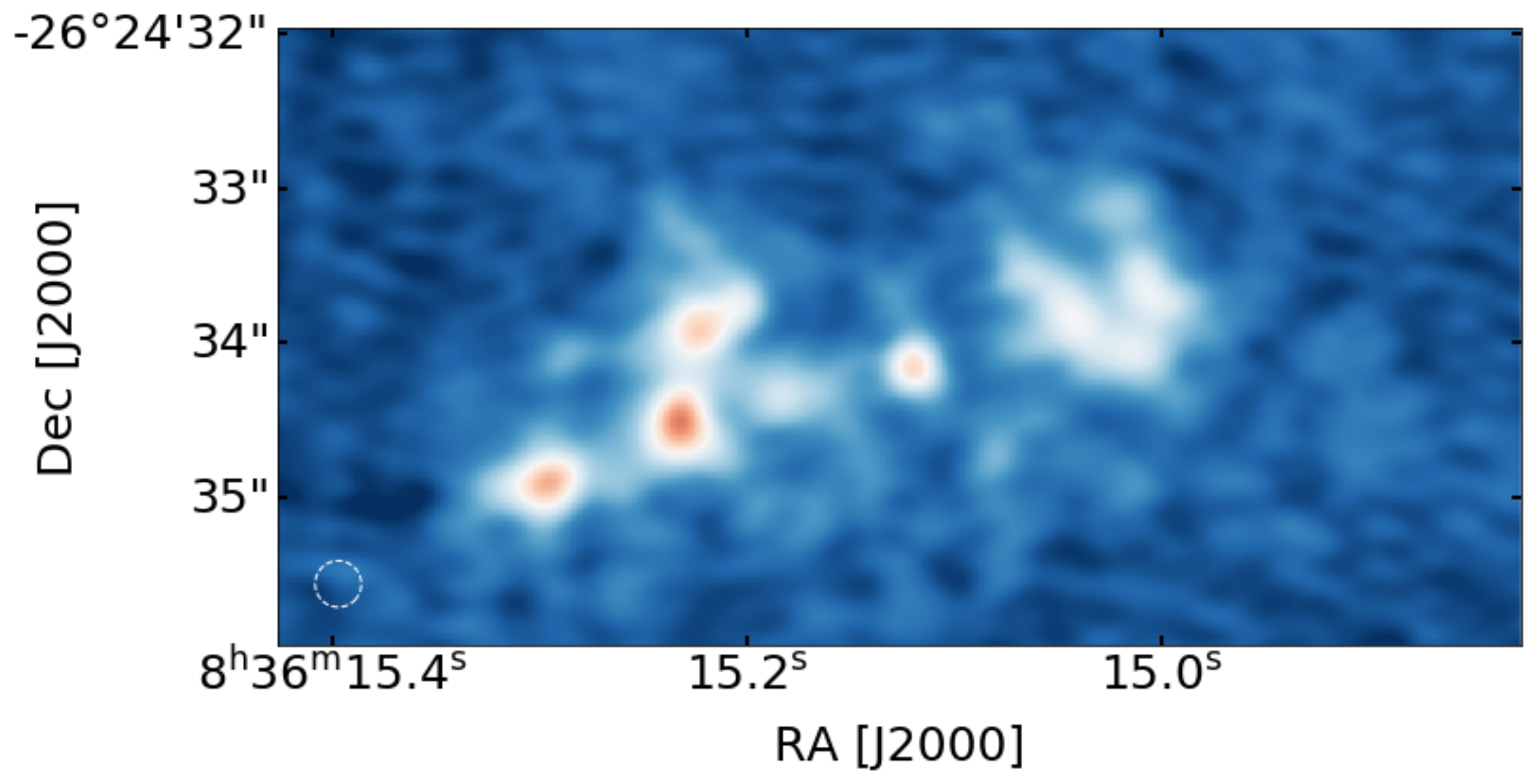}{0.49\textwidth}{(d) 22 GHz}}

\gridline{\fig{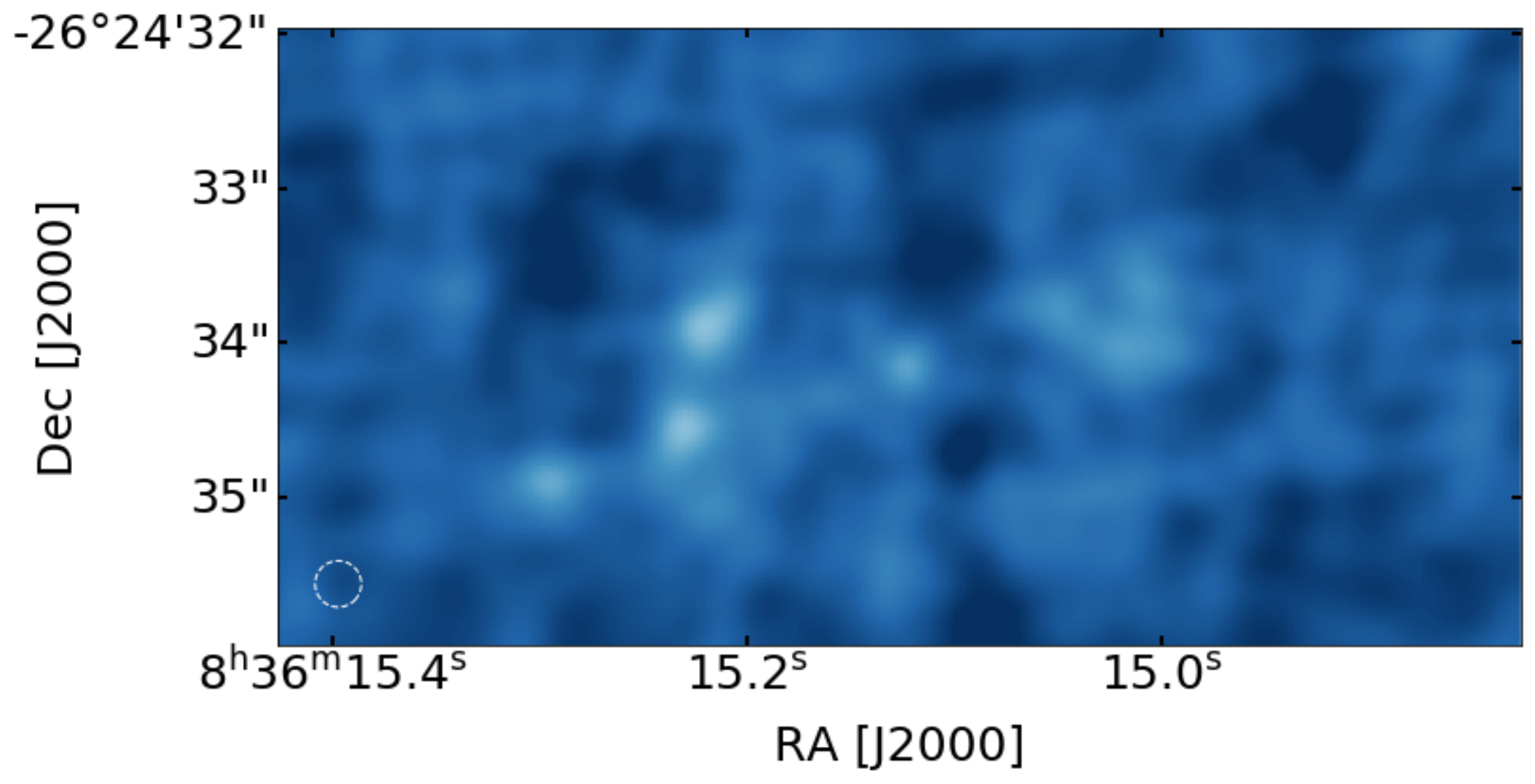}{0.49\textwidth}{(e) 33 GHz}
          \fig{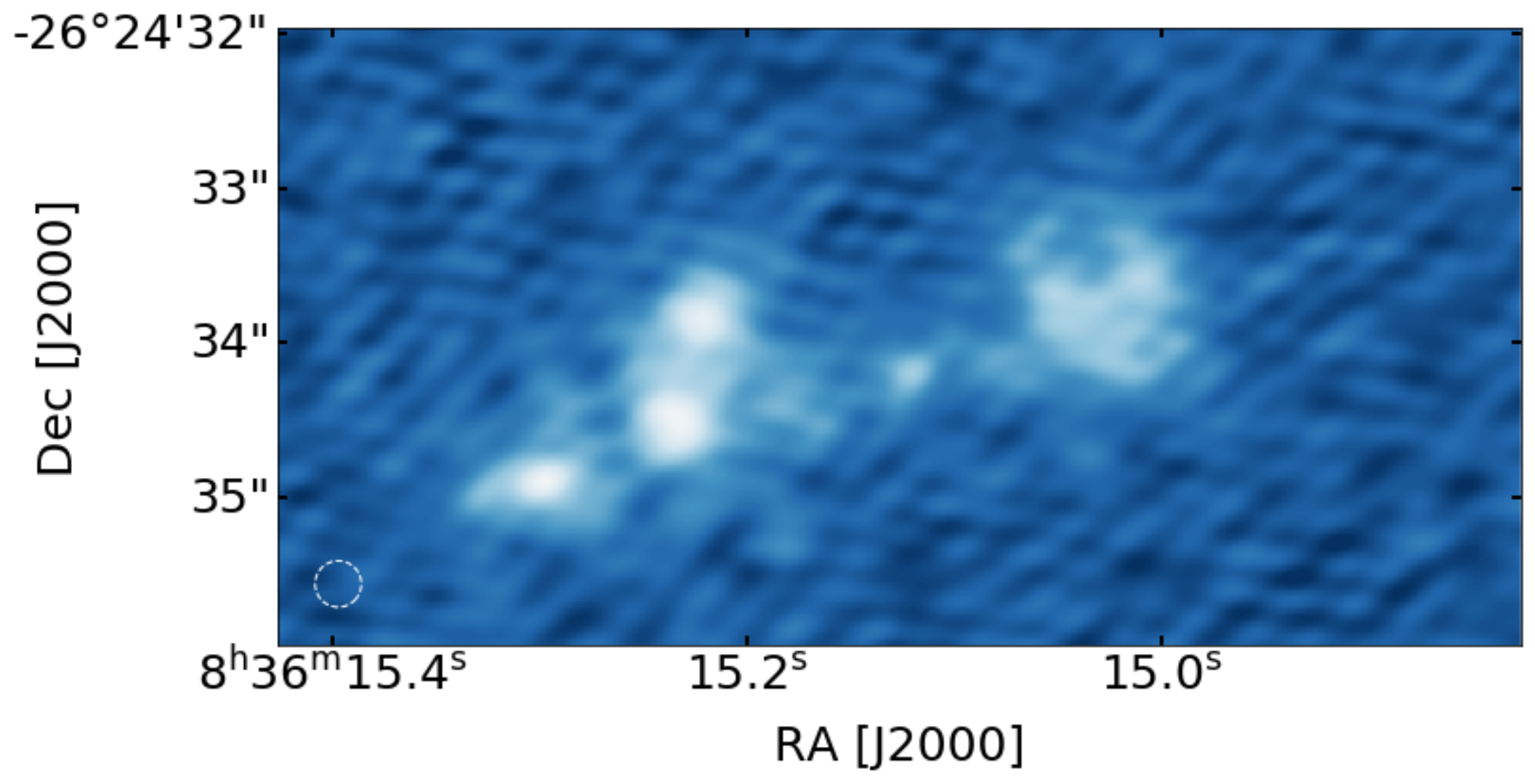}{0.49\textwidth}{(113 GHz)}}

\gridline{\fig{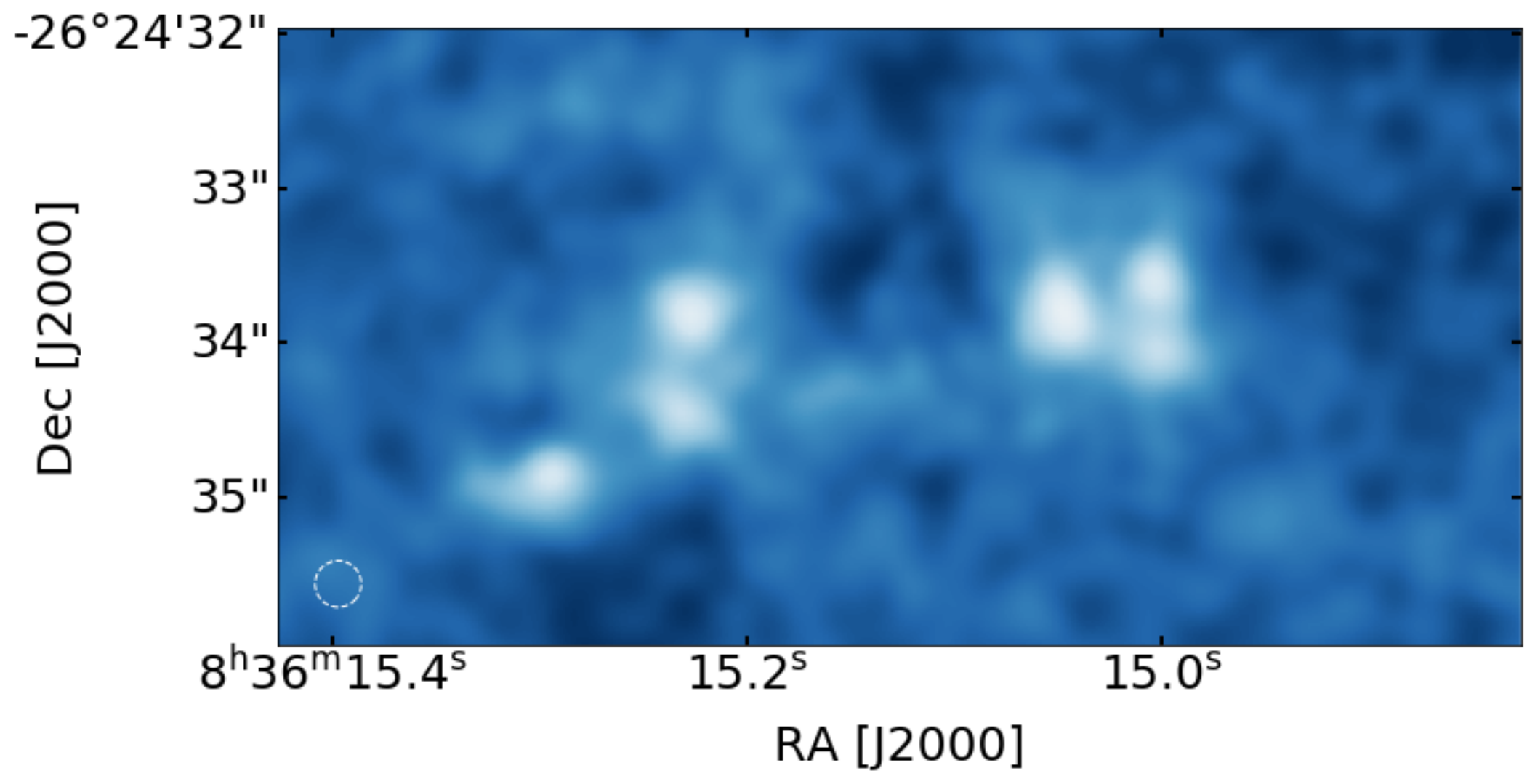}{0.49\textwidth}{f) 250 GHz}
          \fig{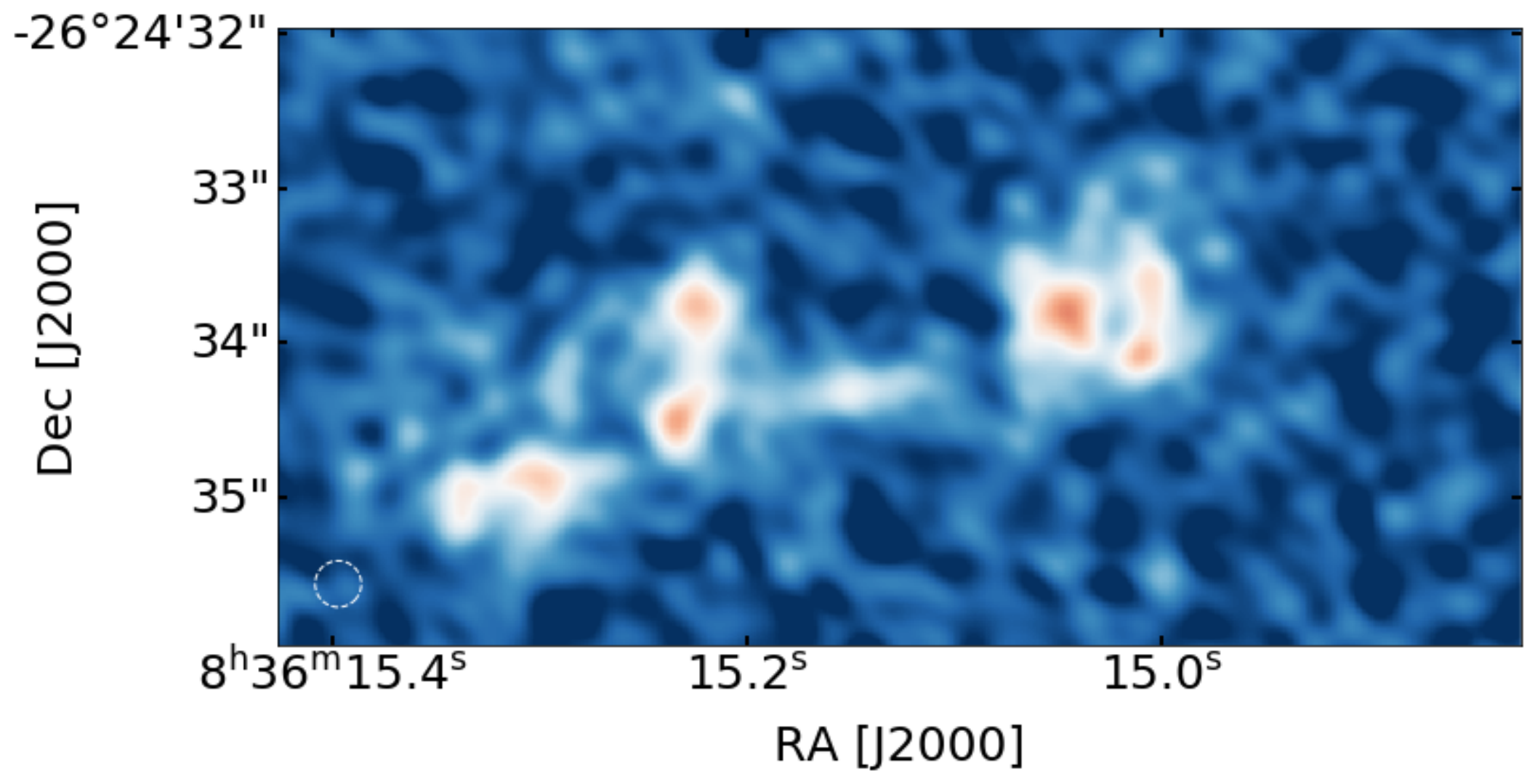}{0.49\textwidth}{(g) 340 GHz}}          
\gridline{\fig{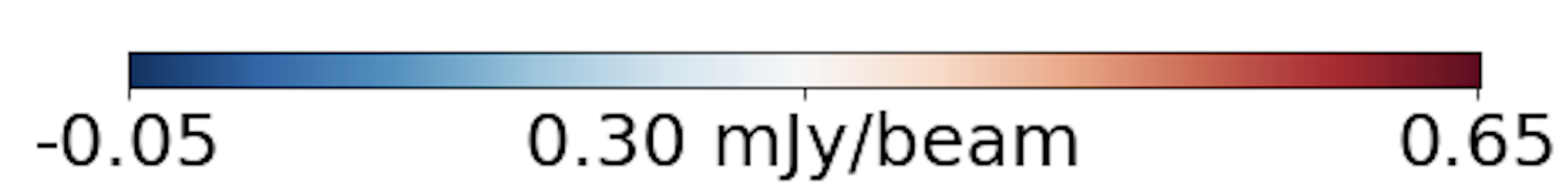}{0.5\textwidth}{}}

\caption{The maps have been convolved to a common beam of 0.3\arcsec{} $\times$ 0.3\arcsec. The beam is shown in the lower left, and all of the maps use the intensity scaling of the convolved 8 GHz map (colorbar).}
      \label{fig:all1}
\end{figure}

Ideally for photometry, the maps across all frequencies would be convolved to a common resolution that is set by the largest resolution available. The largest resolution is the C-band (5 GHz) map at 0.620\arcsec $\times$ 0.248\arcsec. While this resolution is an improvement on previous work in \citet{Johnson:2003}, the majority of the maps have a better (higher) resolution than the older 2004 C- and X-band data. We convolved the maps to a common synthesized beam of 0.3\arcsec{}, which is roughly the synthesized beam at a robust of 0.5, excluding the two historical VLA programs. These maps are shown in Figure \ref{fig:all1}. For C-band, this is a significant super-resolution of the data in the declination direction. The images are {\sc clean}ed deeply, and we estimate from comparing the model and residuals from {\sc clean} that only 5$\%$ of the flux density remains in the residuals. This uncertainly is less than uncertainty due to the flux calibration. For X-band, this enforced restoring beam amounts to beam reshaping and modest super-resolution in the declination direction. Similarly, for the ALMA Band 3 (113 GHz) map, the forced restored image at a 0.3\arcsec{} round beam amounts to moderate super-resolution in the declination direction. We note again that the 33 GHz data suffers from decorrelation; we treat the extracted flux densities from the 33 GHz data as a lower limit.

\citet{Johnson:2003} convolved the maps to a common synthesized beam of 0.95\arcsec{} $\times$ 0.44\arcsec{} for their analysis of five radio knots. With our higher resolution maps, the morphologies of the radio knots are different, and in some cases, we identify multiple components within a knot. Because the photometry apertures used in this work are not identical to those of \citet{Johnson:2003}, we expect that our reported flux densities may not agree with those reported by \citet{Johnson:2003}. However, we expect the differences to be a dependent on the aperture used and not absolute flux calibration. As a check, we convolved the 5, 8, and 15 GHz maps from programs AJ314 and 16B-067 to the synthesized beam from \citet{Johnson:2003} to compare the flux calibrations against their 1995 AK0400 program. We are in possession of the original FITS images used in \citet{Johnson:2003} and are able to make a direct comparison without needing to re-reduce the data from the archive.

We find that for any given size of aperture, our 2004 AJ314 5 GHz map is $\sim$2$\times$ higher in flux density than the 1995 AK0400 5 GHz map, which was previously noted by \citet{Reines:2011}. This is cause for concern, as there is disagreement between the data sets despite the similar observing configurations. Fortunately, both observing programs included a secondary calibrator that can be used as a check on the flux calibration. When calibrated as a science target, the secondary calibrator (J1146+399) in the 1995 AK0400 data has a flux is a factor of 2 lower than the reported flux density in the updated VLA Calibrator Manual. However, the secondary calibrator (J0713+438) in the 2004 AJ314 data agrees with the VLA Calibrator Manual. We re-reduced the 1995 data to verify the calibration and found similar results. It is understood, however, that the VLA calibrator manual may be out of date due to time variability of the flux densities of the calibrators. The flux densities reported in Table \ref{tab:fluxcontour} are not modified, i.e., they are the values extracted from the calibrated 2004 AJ314 data. However to account for the discrepancy between the calibrations, we adjust the flux density and error bars when we model the spectral energy distribution in Section \ref{sec:SED}. We discuss this modification further in that section.

Similarly, the 15 GHz data does not agree between the two data sets. We convolved the 16B-067 15 GHz map to the resolution of the AK0400 map, and for any aperture size we use between the two, the 16B-067 15 GHz map flux density is $\sim$1.4$\times$ lower than that of the AK0400 map. Neither the 15 GHz AK0400 nor the 16B-067 program observed a secondary flux calibrator, so we cannot check the calibration against the calibrator manual. As with the 5 GHz data, we modify the 15 GHz flux densities for the SED analysis in Section \ref{sec:SED}. We did not find similar inconsistence between the 2004 AJ314 8 GHz data and the AK0400 8 GHz map; the flux densities extracted for any given aperture size from both data sets agree within the errors.

\subsubsection{Source Identification and Photometry for Sources Extracted from the Convolved Maps\label{sec:sourceextra}}

\citet{Kobulnicky:1999} and \citet{Johnson:2003} identify five radio knots in their low resolution maps, 0.8\arcsec{} $\times$ 0.4\arcsec{} and 0.95\arcsec{} $\times$ 0.44\arcsec, respectively. At a distance of 9 Mpc, the low resolution maps probe spatial scales of $\sim$35 pc. Our maps convolved to a common synthesized beam of 0.3\arcsec{} (13 pc) are $\sim$3$\times$ higher in resolution, and by inspection of the maps in their native resolutions, we identify additional sources from the aggregate complexes initially identified by \citet{Kobulnicky:1999}. Comparison of our maps and Figure 1 of \citet{Johnson:2003} shows that their Knot 4 is at least two sources, which we label as Knot 4a and 4b. Knots 1 and 2 appear to also have an addition component in their vicinity. We label these three sources as Knot 1a, 1b, and 2; however, it is difficult to disentangle Knot 1b in relation to Knots 1 and 2 of the original maps. We could have just as easily called Knot 1b as Knot 2b and direct comparison between our Knot 2 and the original Knot 2 in \citet{Johnson:2003} needs to be done cautiously. 

For photometry, we stacked the convolved 0.3\arcsec{} maps, which includes all the maps listed in Table \ref{tab:alldataprop}, except for the HST/NICMOS map. The stacked image reinforces persistent sources over the wide range of frequencies, but it likely poorly preserves morphological differences and may only capture the high surface brightness features. We overlaid contours of the stacked image and selected the contour of 20$\times$ the rms (73 $\mu$Jy bm$^{-1}$) of the stacked map. All seven sources are present in the stacked map. In the individual maps, the contour level chosen corresponds to at least a 5$\sigma$ contour, except for the 340 GHz map. In the 340 GHz map, this would be equivalent to a 3$\sigma$ contour. We manually inspected a range of contour levels in the stacked and individual maps; we concluded that a 20$\sigma$ contour best represents the morphology of the sources across all the frequency maps and maximizes the flux included in the analysis while minimizing contamination from noise. The contours have sharp edges in crowded regions because we generously masked sources to not double count extended faint emission in the pixels between the sources. While we employed a mask by manually inspecting the emission in the maps, our choice of pixel assignment in the fainter emission does not significantly change the results. The errors reported on the flux densities are dominated by uncertainties in the flux calibration.

\begin{figure}[htb!]
\centering
\includegraphics[width=0.7\textwidth]{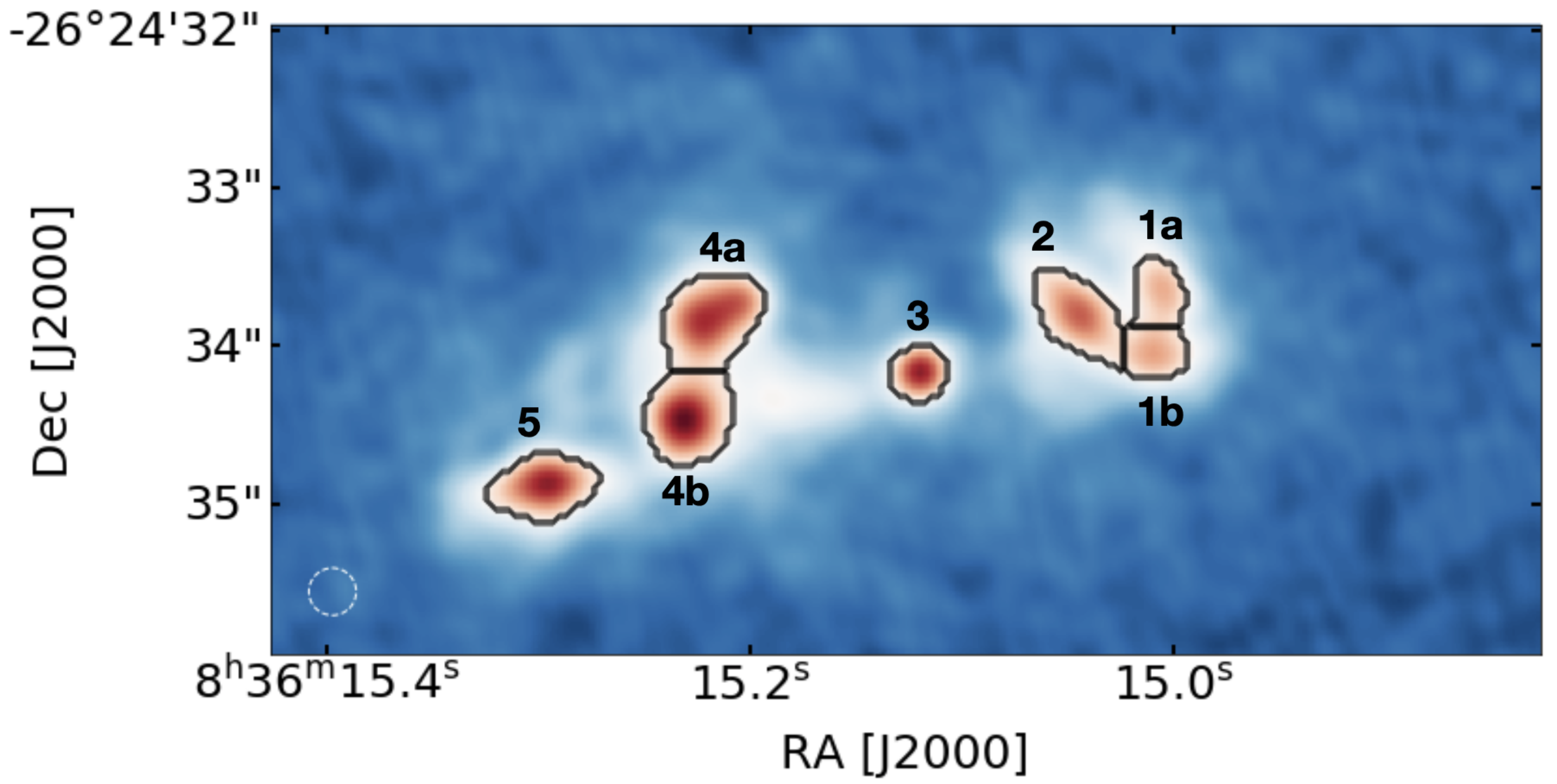}
\caption{Stacked image of all the convolved maps. The extracted sources are outlined in black and represent 20$\times$ the rms of the stacked image.}
\label{fig:stacked}
\end{figure}

Figure \ref{fig:stacked} shows the contours that delineate the extent of the sources in the stacked map. We identify seven sources, whose labels and coordinates are given in Table \ref{tab:coords}. The semi-major and -minor axes in columns 4 and 5 are a 2D Gaussian model of the region extracted by the contour to approximate the size of each source. The errors reported on the flux density follow \citet{Condon:1997} for elliptical Gaussian fits and incorporate the systematic errors from the telescopes. For the ALMA bands, the systematic errors are 10$\%$ \citep{Fomalont:2014}, and for the VLA, 5$\%$ for frequencies $<$ 15 GHz and otherwise 15$\%$\footnote{\url{https://science.nrao.edu/facilities/vla/docs/manuals/oss/performance/fdscale}}. We used the same contour footprint in each map to calculate the integrated flux density in the same manner as CASA\footnote{\url{https://casa.nrao.edu/casadocs-devel/stable/global-task-list/task_imstat/about}}.  To briefly summarize, we sum the intensity across the pixels associated with the source and divide it by the beam area. The beam area is defined as \(\textrm{beam area} = \frac{\pi}{4\ln{2}} \theta_{maj} \theta_{min}, \) where $\theta_{maj}$ and $\theta_{min}$ are the major and minor axes of the beam respectively. For the background subtraction, we masked the sources and calculated the median intensity of the image. We then multiplied the median background intensity by the number of beam areas for the source to find a background flux density. Table \ref{tab:fluxcontour} lists the background subtracted flux densities for extracted sources. 

\begin{table}[htb!]
\centering
\begin{threeparttable}
\caption{Properties of Radio Knots from Photometry\label{tab:coords}}
\begin{tabular}{P{2cm}P{2cm}P{3cm}P{1cm}P{1cm}}
\toprule
\multirow{2}{*}{Knot} & \multirow{2}{*}{RA} & \multirow{2}{*}{decl} & a$_{maj}$\tnote{a} & b$_{min}$ \\ 
 & & & (\arcsec) & (\arcsec) \\
\midrule
1a & 8$^{h}$36$^{m}$15.01$^{s}$ & --26\ddeg24\arcmin33.66\arcsec &  0.23 & 0.17\\ 
1b & 8$^{h}$36$^{m}$15.01$^{s}$ & --26\ddeg24\arcmin34.10\arcsec & 0.22 & 0.17\\ 
2 & 8$^{h}$36$^{m}$15.05$^{s}$ & --26\ddeg24\arcmin33.86\arcsec  & 0.37 & 0.20\\ 
3 & 8$^{h}$36$^{m}$15.12$^{s}$ & --26\ddeg24\arcmin34.16\arcsec & 0.19 & 0.17\\ 
4a & 8$^{h}$36$^{m}$15.22$^{s}$ & --26\ddeg24\arcmin33.86\arcsec & 0.36 & 0.27\\ 
4b & 8$^{h}$36$^{m}$15.24$^{s}$ & --26\ddeg24\arcmin34.50\arcsec & 0.31 & 0.27 \\ 
5 & 8$^{h}$36$^{m}$15.30$^{s}$ & --26\ddeg24\arcmin34.90\arcsec & 0.34 & 0.20\\ 
\bottomrule
\end{tabular}
\begin{tablenotes}
\item[a] Approximate semi-major (a$_{maj}$) and semi-minor (b$_{min}$) axes from the contoured regions in Figure \ref{fig:stacked}.
\end{tablenotes}
\end{threeparttable}
\end{table}

\begin{table}[htb!]
\begin{threeparttable}
\centering
\caption{Background Subtracted Flux Densities and Uncertainties of Radio Knots.\label{tab:fluxcontour}}
\setlength\tabcolsep{3pt}
\begin{tabular}{P{1.3cm}P{1.7cm}P{1.7cm}P{1.7cm}P{1.7cm}P{1.7cm}P{1.7cm}P{1.6cm}}
\toprule
Freq & \multicolumn{7}{c}{Knot} \\\cmidrule{2-8}
(GHz) & 1a & 1b & 2 &  3 & 4a & 4b & 5 \\ 
\hline
5 & 0.27 $\pm$  0.03 & 0.38 $\pm$  0.03 & 0.74 $\pm$  0.06 & 0.66 $\pm$  0.04 & 1.35 $\pm$  0.09 & 1.07 $\pm$  0.07 & 0.89 $\pm$  0.06 \\ 
8 & 0.31 $\pm$  0.05 & 0.28 $\pm$  0.05 & 0.61 $\pm$  0.10 & 0.46 $\pm$  0.07 & 1.07 $\pm$  0.17 & 0.86 $\pm$  0.13 & 0.75 $\pm$  0.12 \\ 
15 & 0.23 $\pm$  0.04 & 0.25 $\pm$  0.04 & 0.38 $\pm$  0.08 & 0.24 $\pm$  0.03 & 0.57 $\pm$  0.10 & 0.71 $\pm$  0.08 & 0.42 $\pm$  0.07 \\ 
22 & 0.29 $\pm$  0.06 & 0.27 $\pm$  0.06 & 0.54 $\pm$  0.11 & 0.26 $\pm$  0.05 & 0.77 $\pm$  0.14 & 0.81 $\pm$  0.14 & 0.61 $\pm$  0.11 \\ 
33 & 0.10 $\pm$  0.04 & 0.11 $\pm$  0.04 & 0.16 $\pm$  0.06 & 0.08 $\pm$  0.02 & 0.28 $\pm$  0.07 & 0.25 $\pm$  0.06 & 0.16 $\pm$  0.05 \\ 
113 & 0.19 $\pm$  0.04 & 0.17 $\pm$  0.04 & 0.38 $\pm$  0.08 & 0.13 $\pm$  0.03 & 0.55 $\pm$  0.09 & 0.57 $\pm$  0.09 & 0.40 $\pm$  0.06\\ 
258 & 0.24 $\pm$  0.05 & 0.22 $\pm$  0.05 & 0.47 $\pm$  0.09 & 0.07 $\pm$  0.03 & 0.51 $\pm$  0.10 & 0.44 $\pm$  0.10 & 0.37 $\pm$  0.08 \\ 
340 & 0.36 $\pm$  0.09 & 0.37 $\pm$  0.08 & 0.77 $\pm$  0.16 & 0.13 $\pm$  0.05 & 0.75 $\pm$  0.17 & 0.71 $\pm$  0.15 & 0.62 $\pm$  0.15 \\ 
\midrule
\midrule
 $\lambda$1.876\tnote{b} & \multirow{2}{*}{0.31 $\pm$ 0.24} & \multirow{2}{*}{1.07 $\pm$ 0.22} & \multirow{2}{*}{2.42 $\pm$ 0.26} & \multirow{2}{*}{0.20 $\pm$ 0.13} & \multirow{2}{*}{3.12 $\pm$ 0.26}  & \multirow{2}{*}{3.08 $\pm$ 0.25} & \multirow{2}{*}{0.55 $\pm$ 0.21}\\
 $\mu$m & \\
\hline
\end{tabular}
\begin{tablenotes}
\item[a] All numerical entries given in mJy except for column one.
\item[b] Flux density in Pa$\alpha$ map for each knot.
\end{tablenotes}
\end{threeparttable}
\end{table}

\subsubsection{Photometry in \palpha{} Map \label{sec:photnicmos}}

\begin{figure}[htb!]
\centering
\gridline{\fig{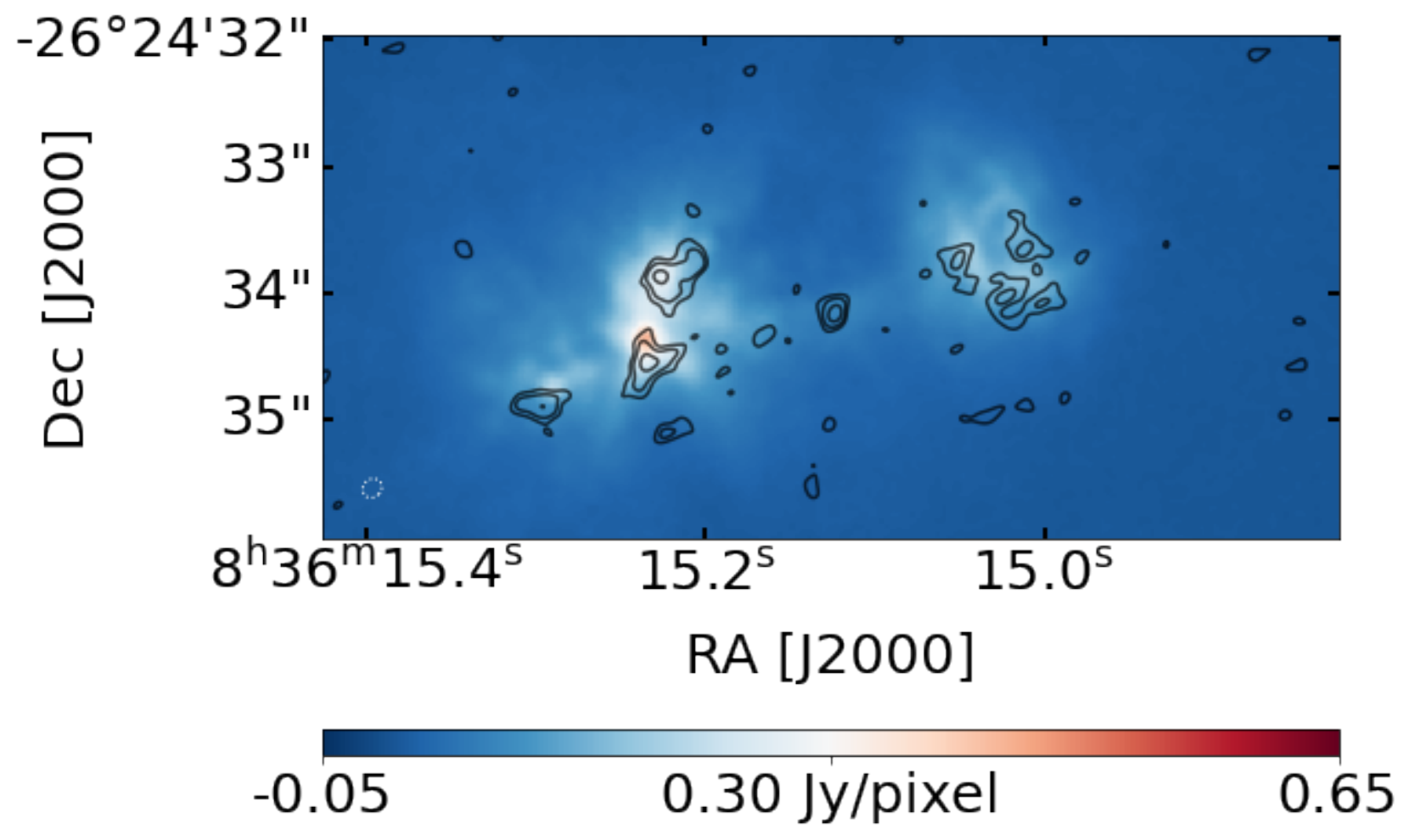}{0.49\textwidth}{(a)}
          \fig{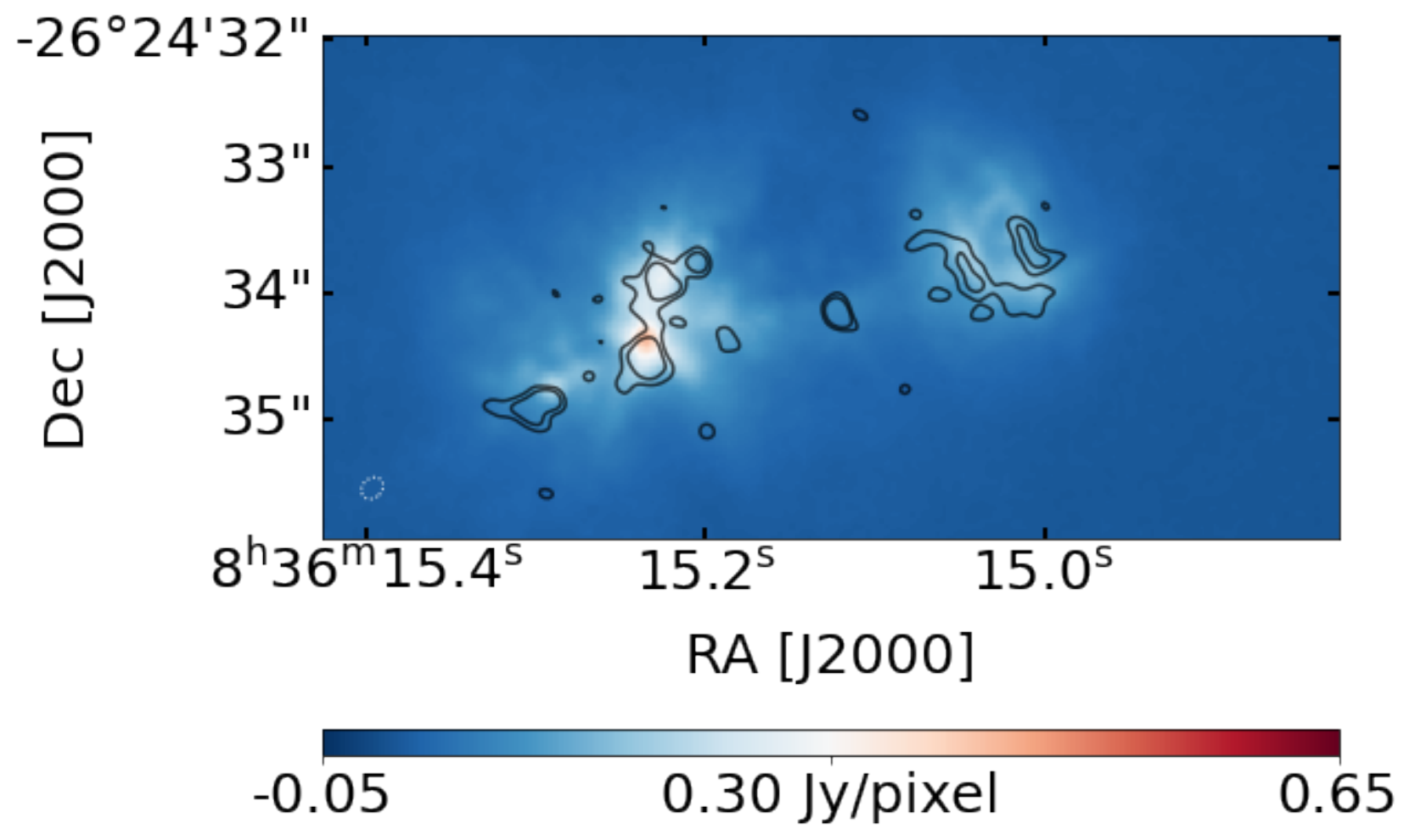}{0.49\textwidth}{(b)}}

\caption{Continuum subtracted HST/NICMOS \palpha{} map with (\textit{a}) contours at 2, 3, $\&$ 5$\sigma$ of the unconvolved 33 GHz map and (\textit{b}) contours at 3 $\&$ 5$\sigma$ of the unconvolved 22 GHz map. In both images, the beam of the radio map is the dotted white circle in the lower left. The raster scale is normalized to the peak of convolved 8 GHz maps, as in Figure \ref{fig:all1}.} 
\label{fig:nicmos}
\end{figure}

Figure \ref{fig:nicmos} is a continuum subtracted HST/NICMOS \palpha{} map; the raster scale is normalized to the peak in the convolved 8 GHz map for ease of comparison between this figure and Figure \ref{fig:all1}. We initially extracted the flux density for each source using the contours from the stacked image, similar to the photometry in the other maps. With these apertures, the extracted regions for Knots 4a, 4b, and 5 include features in the \palpha{} map that may not actually be associated with the radio knots. Of particular note is the bright \palpha{} feature located near Knot 4b. This compact feature appears to have optical counterparts, such as those seen in Figure 3 of \citet{Cabanac:2005}, but is offset from the radio peak. Inclusion of this compact feature alone significantly boosts the flux density by a factor of $\sim$2.6 for Knot 4b, which has undesirable consequences in our calculation of the extinction in Section \ref{sec:extinction}: namely, it produces a negative extinction value for this line of sight.

To avoid unphysical extinction values, we extracted the flux density for a region that corresponds to the half light radius of each source in the 33 GHz map, which roughly corresponds to a 2$\sigma$ contour. The apertures are approximately 30$\%$ smaller with this treatment. The main motivation in using the half light radius is the morphology of the 33 GHz data for Knots 4a and 5 because the bright features in the \palpha{} map are not included in the half light contour. We believe that examination of Figure \ref{fig:nicmos} supports this decision, but the best method of extraction likely lies between these two approaches. We compared the results of the methods as a sanity check; the change in aperture approximately acts like a scale factor to the calculated values and does not ultimately change our interpretation of the results. We also performed similar photometry using the unconvolved 22 GHz maps to set the contours (Figure \ref{fig:nicmos} right) because of the data reduction concerns in the 33 GHz data. The extracted flux density using the 22 GHz contours have a factor 2 difference for Knots 2 and 4a; the other sources are consistent within the errors. The final interpretation remains unchanged, so we retain the use of the 33 GHz contours in the subsequent analysis. We make note of this approach in latter sections when appropriate to remind the reader that photometry of extended sources is an art.

Regardless of the method used, we place an annulus around the source with an inner and outer radii of 1.5 and 2 times the size of the source for the background estimate. The exception is Knot 4b, for which we used an annulus with an inner and outer radii of 2 and 3 times the size of the source to not include the nearby bright, compact feature in the background subtraction. Table \ref{tab:fluxcontour} gives the background subtracted flux density of the sources using the half light contour approach.

\subsection{Remarks on Photometry}
Robust photometry of extended sources is challenging, particularly in crowded fields, but it is crucial to recover low surface brightness features to fully characterize the extended sources. We investigated two additional methods of source identification and source extraction beyond what we have discussed above. In addition to the contour method, we used the CASA task {\sc imfit} to extract sources from manually drawn polygonal regions. We also utilized an automated source identification program in the Python package \textit{Photoutils}\footnote{\url{https://photutils.readthedocs.io/en/stable/index.html}} v0.7.2 \citep{Bradley:2019} to identify sources. The Photoutils $detect\_sources$ module detects both point-like and extended sources and then deblends overlapping sources with the $deblend\_sources$ module. The output from the $detect\_sources$ module was then used as initial guesses for a 2D Gaussian fitting routine to extract the best fit 2D Gaussian model for the source at each frequency.

The extraction of extended sources is complicated and particularly sensitive to the size of aperture used, regardless of the method. Accounting for the extended emission and disentangling crowded fields is difficult to achieve without some level of subjectivity. We compared the results of these routines, and we determined that the extraction based on a contour level was the least dependent on user input, once the initial level was specified. This method best mitigated the subjectivity in the uncertainties inherent to extracting extended structures.

\section{Results and Discussion}

\subsection{Spectral Energy Distributions from 0.9 mm to 6 cm\label{sec:SED}}

UD \HII regions at radio wavelengths are described by a thermal model with a turnover frequency $\gtrsim$ 5 GHz \citep{Wood:1989,Johnson:2003}, but the presence of SNRs in the vicinity may introduce a non-thermal component. For completeness, we allow for this possibility by including a non-thermal component given that low-level non-thermal emission could be present. The ALMA data will probe the dust emission associated with the sources, so we include a dust component as well to disentangle the non-thermal, thermal, and dust emission, as only the thermal emission will provide an estimate of the rate of ionizing photons and thus the stellar content of the sources. \textit{A priori}, we expect each component to follow a power law. The free-free emission is likely optically thin with a spectral index of $\alpha_{ff}$ = --0.1 above 5 GHz \citep{Allen:1998, Kobulnicky:1999}. We adopt an optically thin dust component of $\alpha_{dust}$ = 4 \citep{Scoville:2014}. SNRs can have $\alpha_{nth} \lesssim$ --0.5 or steeper, depending on the age of SNR \citep{Weiler:1986,Bell:2011,Dubner:2015}.  We compared the models with $\alpha_{nth}$ = --0.7, which is a typical synchrotron spectral index, and $\alpha_{nth}$ = --0.5 and found that the reduced chi-squared, $\chi_{\nu}^2$, of $\alpha_{nth}$ = --0.7 models are slightly better (i.e., closer to 1) than the models with --0.5, thus we only show the results of $\alpha_{nth}$ = --0.7 in the sections below. The exception is Knot 3; for both the 2 and 3 component models, $\alpha_{nth}$ = --0.5 is a better fit to the data.

\begin{figure}[htb!]
 \centering
\gridline{\fig{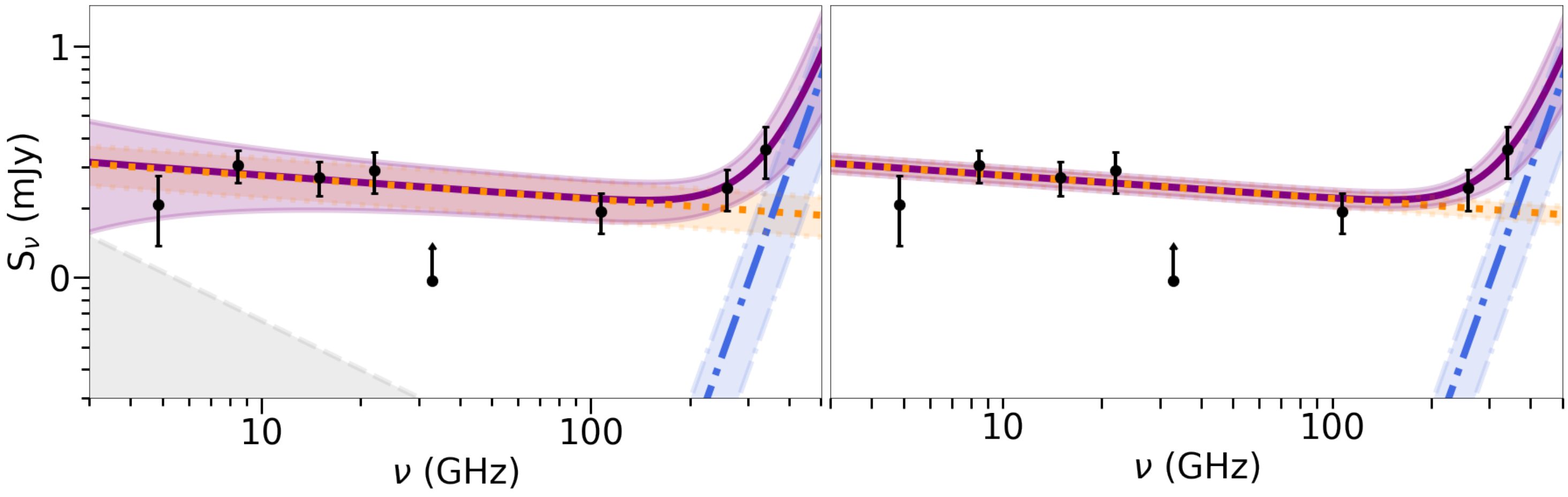}{0.8\textwidth}{(a) Knot 1a}}

\gridline{\fig{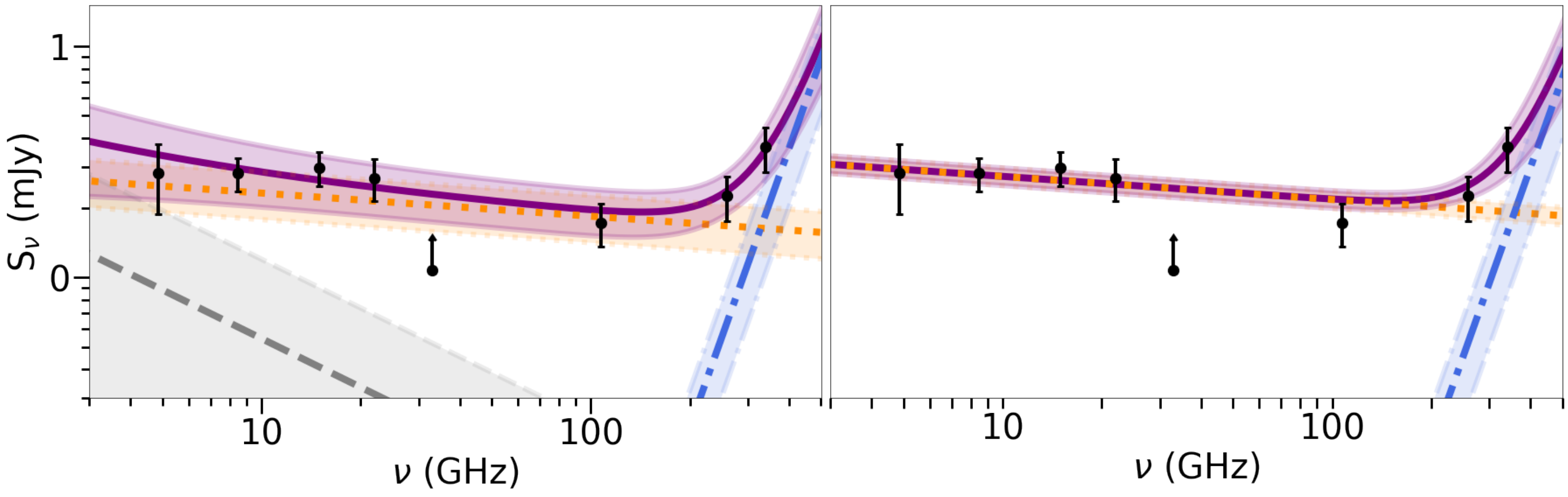}{0.8\textwidth}{(b) Knot 1b}}

\gridline{\fig{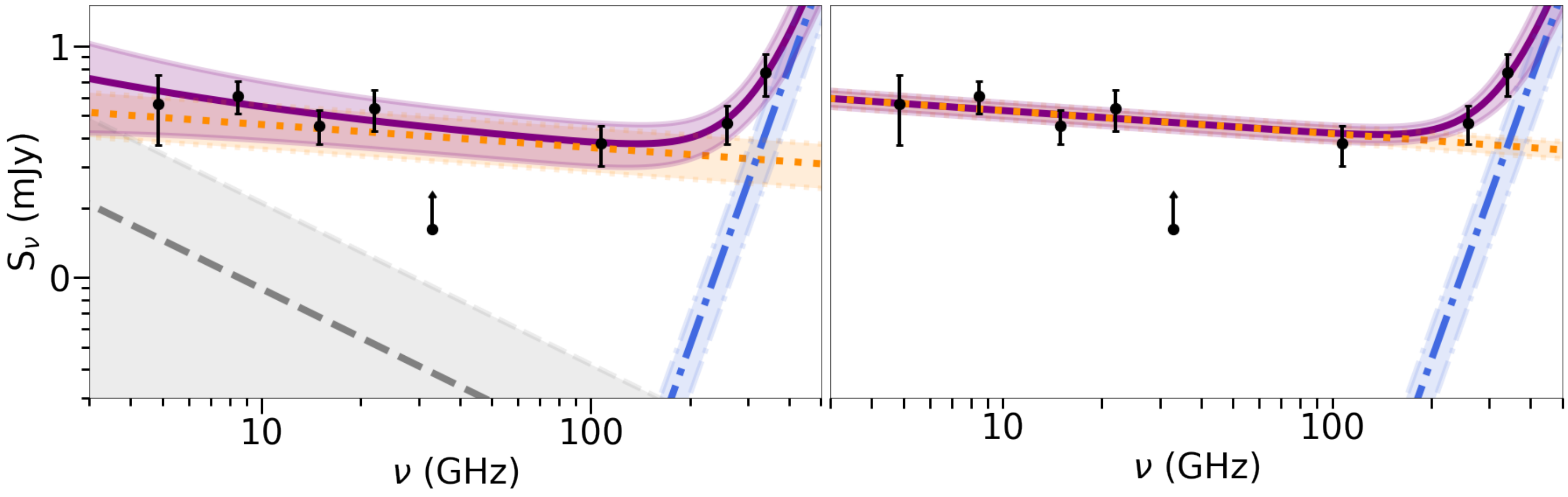}{0.8\textwidth}{(c) Knot 2}}
\caption{Plots of flux density and uncertainties for Knots 1a, 1b, and 2. The left column shows the best fit SED model (solid purple) with free-free (dotted orange), non-thermal (dashed gray), and dust (dotdashed blue) power-law components. The right column shows the best fit SED for the two component fit. The shaded regions associated with each line represent the 1$\sigma$ errors on the fit parameters. Though displayed, the 33 GHz flux density was excluded from the fit as it is a lower limit. The 5 and 15 GHz flux densities were modified from the values in Table \ref{tab:fluxcontour} as discussed in the text.}
 \label{fig:contour1}
\end{figure}

\begin{figure}[htb!]
 \centering
 
 \gridline{\fig{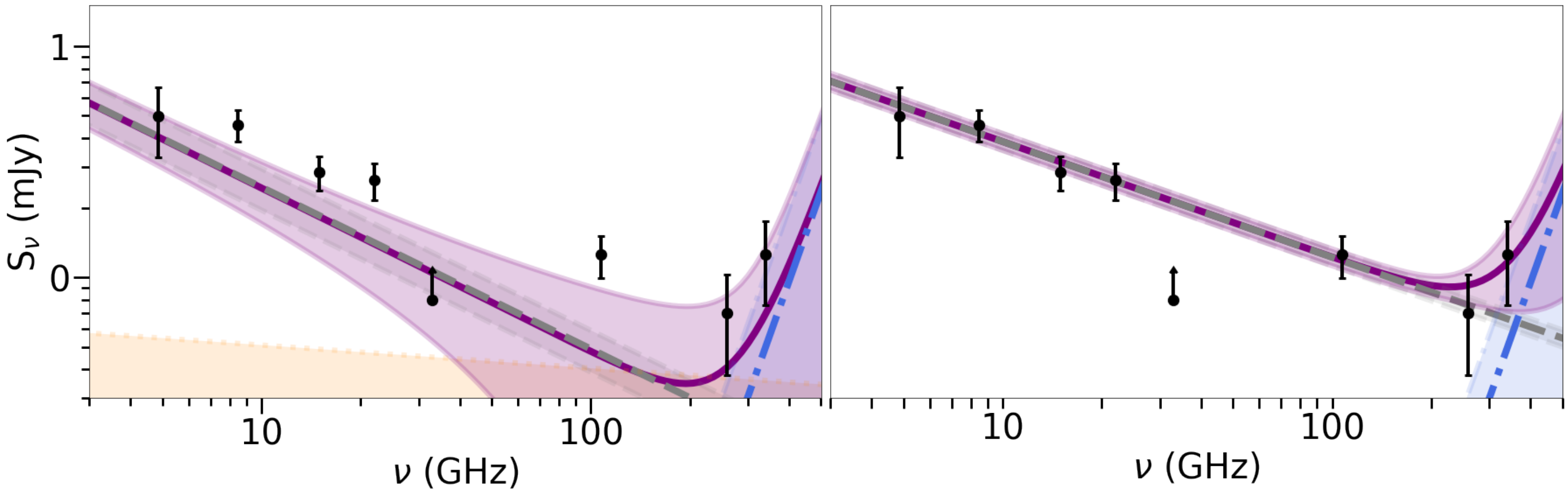}{0.8\textwidth}{(a) Knot 3}}

\gridline{\fig{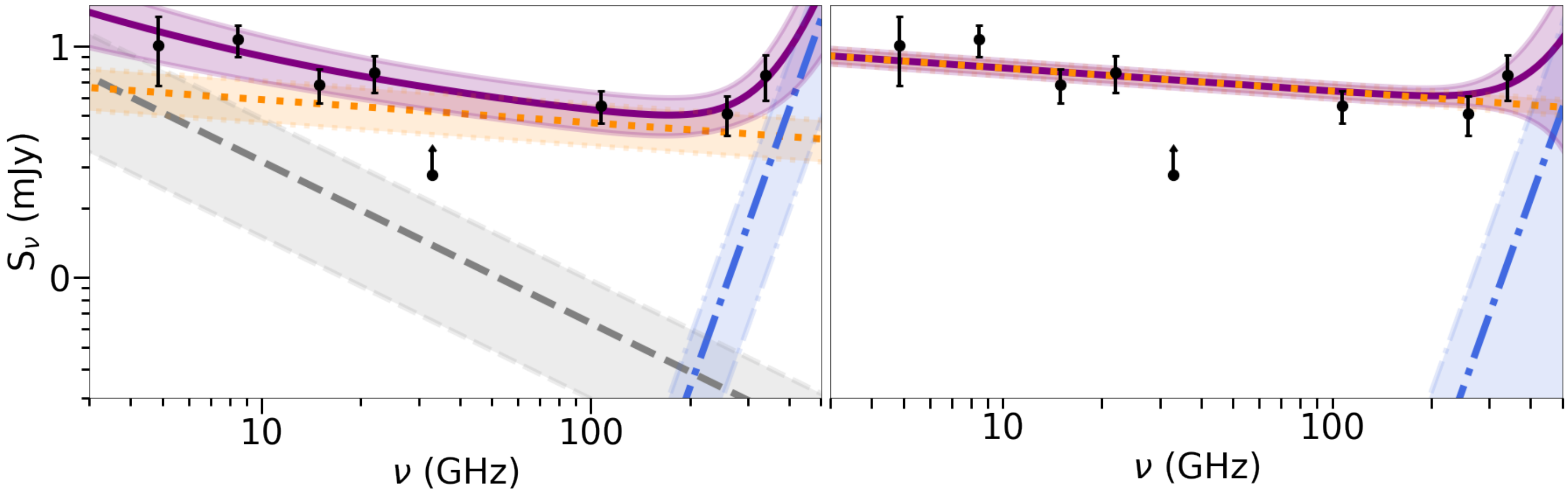}{0.8\textwidth}{(b) Knot 4a}}

\gridline{\fig{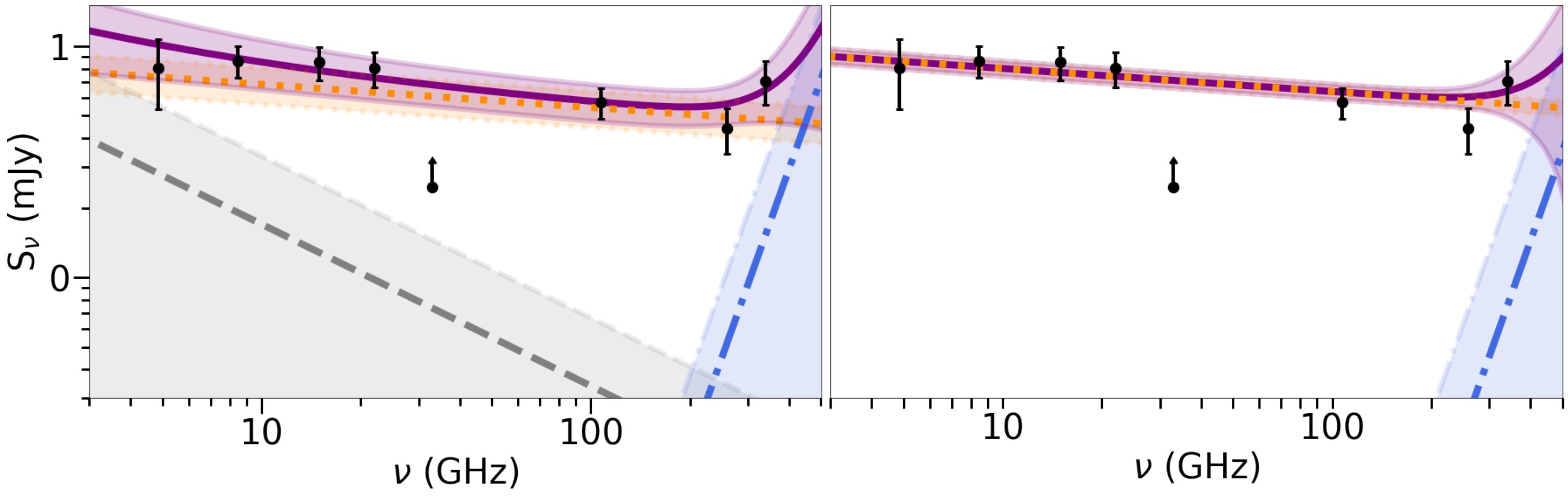}{0.8\textwidth}{(c) Knot 4b}}

\gridline{\fig{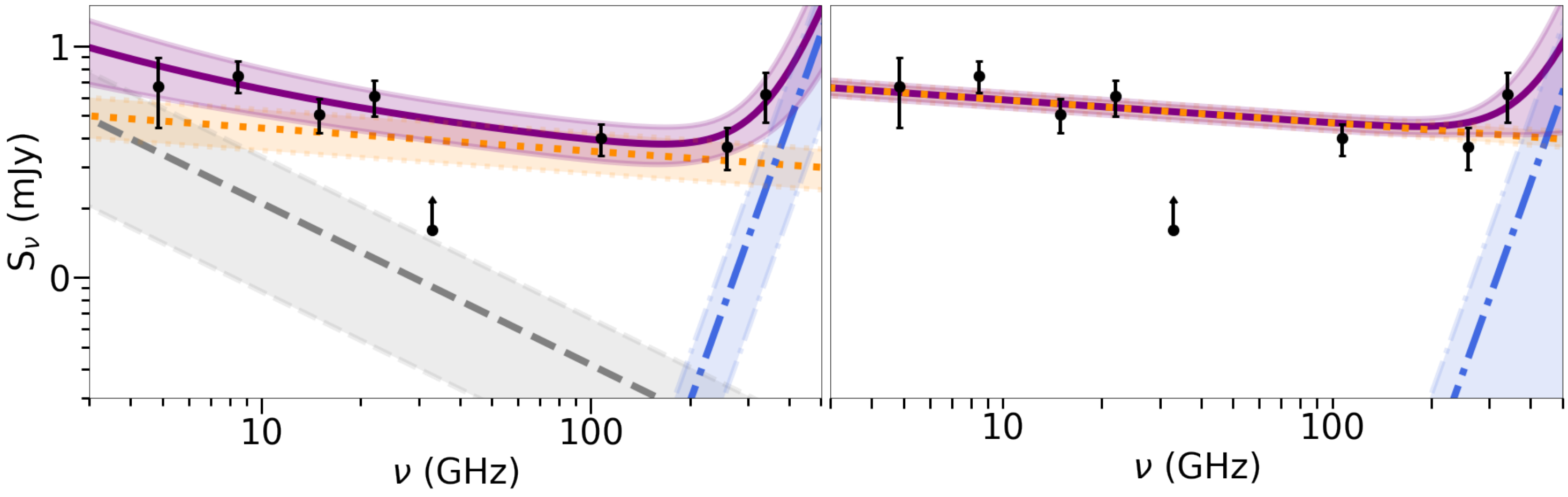}{0.8\textwidth}{(d) Knot 5}}

\caption{Similar to Figure \ref{fig:contour1} but for Knots 3, 4a, 4b, and 5. The two component fit for Knot 3 is non-thermal+dust, and both the 3 and 2 components fits for Knot 3 use $\alpha_{nth}$ = --0.5.}
 \label{fig:contour2}
\end{figure}

We fix the spectral indices to the values listed above, and the model describing the 3 components is
 \begin{equation}
 \log{} S_{model} = \log{ (S_{0,nth}\;\nu^{\alpha_{nth}} +S_{0,ff}\; \nu^{\alpha_{ff}}+S_{0,dust}\;\nu^{\alpha_{dust}}  )}, 
 \label{eq:SED}
 \end{equation}
 where the subscripts \textit{nth}, $ff$, and $dust$ denote non-thermal, free-free, and dust, respectively. We perform a $\chi^2$ minimization for the three free parameters $S_{0,nth}$, $S_{0,ff}$, and $S_{0,dust}$. Table \ref{tab:chicontour} gives the fit parameters and $\chi_{\nu}^2$ of the models, and the left column of Figures \ref{fig:contour1} and \ref{fig:contour2} shows the SEDs with the best fit parameters for the 3 component model from Table \ref{tab:chicontour}. 
 
Due to the short lifespans of UD \HII regions, we do not expect even the most massive of stars associated with the UD \HII region to have undergone a supernova event yet. A two component fit, with just free-free and dust emission, may be more appropriate than one that includes a non-thermal one.  Equation (\ref{eq:SED}) is easily modified for just two terms to model this.  Table \ref{tab:chicontour} also gives the fit parameters for the 2 component model, and the right column of Figures \ref{fig:contour1} and \ref{fig:contour2} shows the corresponding SED plots. 

There are two important notes about the SED models. First, the 33 GHz flux density was excluded from both the two and three component models as it is a lower limit. Second, the 5 and 15 GHz flux densities and errors were modified for the SED models to account for the calibration discrepancies between the data in this study and that of AK0400 (see discussion in Section \ref{sec:photom}). Because the original flux densities reported in \citet{Johnson:2003} employed different aperture sizes than ours here and our 3$\times$ higher angular resolution maps identify multiple components to some sources, we cannot simply include the flux densities from \citet{Johnson:2003} for these sources into the SED models. Instead, we compared the flux densities between the AK0400 maps and ours, when convolved to a common beam, to implement a suitable scaling factor to apply to our flux densities. At 5 GHz, our maps have flux densities $\sim$2$\times$ higher than the AK0400 maps, and at 15 GHz, the 16B-076 data have flux densities $\sim$1.4$\times$ lower than the AK0400 data. The modified flux density at 5 GHz is the midpoint between the flux density in Table \ref{tab:fluxcontour} and half that value, and the error is the distance between the midpoint and the 5 GHz flux density. Similarly, the modified flux density at 15 GHz is the midpoint between the flux density in Table \ref{tab:fluxcontour} at 15 GHz and 1.4$\times$ that value; the errors are the distance between those two values. By modifying the data for the SED models in this way, we can incorporate the discrepancies between the two data sets.

\begin{table}[htb!]
\setlength\tabcolsep{4pt}
\begin{threeparttable}
\centering
\caption{Best fit Parameters from SED Models \label{tab:chicontour}}
\begin{tabular}{P{1cm}P{1cm}P{1cm}P{1cm}P{1cm}P{1cm}P{1.5cm}P{1cm}P{2cm}}
\toprule
\multirow{2}{*}{Knot} & \multirow{2}{*}{$\alpha_{nth}$} & S$_{0,nth}$ & \multirow{2}{*}{$\alpha_{ff}$} & S$_{0,ff}$ & \multirow{2}{*}{$\alpha_{dust}$} & S$_{0,dust}$ & \multirow{2}{*}{$\chi_{\nu}^2$} & No. of \\ 
 & & (mJy) & & (mJy) & & (mJy) & & Components \\ 
 \midrule
\multirow{2}{*}{1a} & -0.7 & 0.01 & -0.1 & 0.35 & 4 & 1.2e-11 & 2.3 & 3\\
 &\mytablefill & \mytablefill & -0.1 & 0.35 & 4 & 1.2e-11 & 1.8 & 2 \\ 
 \hline
\multirow{2}{*}{1b} & -0.7 & 0.27 & -0.1 & 0.29 & 4 & 1.4e-11 & 3.9 & 3 \\ 
 &\mytablefill & \mytablefill & -0.1 & 0.34 & 4 & 1.2e-11 & 2.2 & 2 \\ 
 \hline
\multirow{2}{*}{2} & -0.7 & 0.45 & -0.1 & 0.58 & 4 & 3.2e-11 & 5.2 & 3 \\
 &\mytablefill & \mytablefill & -0.1 & 0.66 & 4 & 2.7e-11 & 3.1 & 2 \\ 
 \hline
\multirow{2}{*}{3} & -0.5 & 1.23 & -0.1 & 0.0 & 4 & 3.8e-12 & 0.9 & 3 \\ 
 & -0.5 & 1.22 &\mytablefill & \mytablefill & 4 & 3.3e-12 & 0.7 & 2 \\ 
 \hline
\multirow{2}{*}{4a} & -0.7 & 1.59 & -0.1 & 0.74 & 4 & 2.1e-11 & 2.0 & 3 \\ 
 &\mytablefill & \mytablefill & -0.1 & 0.88 & 4 & 8.5e-12 & 1.3 & 2 \\
 \hline
\multirow{2}{*}{4b}& -0.7 & 0.85 & -0.1 & 0.86 & 4 & 1.2e-11 & 1.1 & 3 \\
 &\mytablefill & \mytablefill & -0.1 & 1.01 & 4 & 5.6e-12 & 0.8 & 2 \\
 \hline
\multirow{2}{*}{5}& -0.7 & 1.05 & -0.1 & 0.56 & 4 & 1.8e-11 & 2.3 & 3 \\ 
 &\mytablefill & \mytablefill & -0.1 & 0.74 & 4 & 9.9e-12 & 1.4 & 2 \\ 
\bottomrule
\end{tabular}
\end{threeparttable}
\end{table}

\citet{Johnson:2003} fit a purely thermal component with a turnover at $\gtrsim$ 5 GHz for all of the radio knots except Knot 3, which hosts the AGN. As noted in Section \ref{sec:photom}, there are inconsistencies between the flux calibrations between the AJ314 program and the one used in their study, so we have set the uncertainties at 5 GHz accordingly to reflect this discrepancy. With our modified (larger) uncertainties at 5 and 15 GHz, nearly all of these knots could be consistent with a turnover $\gtrsim$ 5 GHz. We have not modeled it because the 5 GHz data from AJ314 without this modification of the uncertainties are not consistent with such a turnover. Upon first inspection of the unmodified 5 GHz flux densities, it appeared as if a non-thermal component would be necessary to characterize the SED; however, we cannot be sure the AJ314 calibration is accurate. New observations at 5 GHz and S-band (2 -- 4 GHz) would clarify if there is a turnover consistent with typical UD \HII region SEDs or whether a non-thermal component is indeed real. If these are truly UD \HII regions, we would not expect a significant non-thermal component. Nevertheless, here we consider possible explanations for a non-thermal component in the SED model.
\begin{enumerate}
\item If stars within a UD \HII region are not formed quasi-instantaneously, there could be more evolved stars that have undergone a  supernova event within them that are not taken into account in a simple formation scenario.
\item While natal SSCs are not expected to yet host supernova events given their extremely young ages, evolved stars in nearby SSCs may have done so.  A small non-thermal component may be residual, non-subtracted background from these more evolved SSCs in the vicinity. He 2-10 has a number of optically identified adolescent SSCs, which could be providing the contaminating non-thermal emission, e.g., Knot 4a, which is known to host a SNR \citep{Reines:2012}.
\item Young stellar objects can be associated with jets, which could be sources of non-thermal emission \citep{Garay:2003,Purser:2016}. \citet{Purser:2016} find 22 ionized jets and jet candidates associated with massive star formation. They suggest that a jet may persist for a short time after an \HII region has formed because they associate some of the ionized jets in their sample with hyper compact, ultra compact, and normal \HII regions. \citet{Cecere:2016} report that there is non-negligible synchrotron emission associated with proto-stellar jets in their models. If there is a population of young stellar objects with jets in these UD \HII regions, this may be a source of non-thermal emission.
\end{enumerate}

The 2 and 3 component SED models have different degrees of freedom, so we cannot directly compare the $\chi_{\nu}^2$ values to determine which model is favored.  For this work, we visually inspect the models and leverage the available knowledge at other wavelengths to determine which model and interpretation is more appropriate for each knot. A visual comparison of the two and three component models suggests that Knots 1a, 1b, 2, and 5 may be consistent with only thermal+dust components. At shorter wavelengths, \citet{Cabanac:2005} do not find optical sources associated with these knots, and from CO(2-1) observations,  \citet{Johnson:2018} report that the knots are associated with molecular material. Given these results, we believe that Knots 1a, 1b, 2, and 5 should retain the classification of UD \HII regions, agreeing with the conclusions of \citet{Johnson:2003} and \citet{Cabanac:2005}. For these sources, we use the 2 component SED model, i.e., one without a non-thermal component, in the remaining sections of this work.

Knot 3, which hosts the central AGN of He 2-10, is not the primary focus of this work because it is not a SSC. For completeness, we have included it in the SED models because it does serve as a sanity check and might be useful to other research programs. Knot 3 does have a stronger non-thermal component than the other knots, which is to be expected. The 3 component SED models are not well constrained, likely because of the adjustment to the 5 GHz flux density and uncertainties. The 2 component fit has non-thermal+dust components, though we did fit thermal+dust component for completeness only to discarded it given the nature of the source and the poor fit.

Knots 4a and 4b originally were identified as one source, Knot 4, in \citet{Kobulnicky:1999}. Knot 4a and 4b are associated with strong optical and IR emission in the maps of \citet{Cabanac:2005} but is depleted in CO(2-1) compared to the other regions \citep{Johnson:2018}. These knots may be a combination of normal \HII regions and SNRs (point 1 above), as suggested by \citet{Cabanac:2005}. Alternatively, there are bright and optically visible SSCs located near Knots 4a and 4b, and VLBI observations from \citet{Reines:2012} report a SNR associated with the optical SSC near to Knot 4a, which could be contaminating the photometry results (point 2).  We cannot discriminate between these two scenarios without first resolving the calibration concerns at 5, 15, and 33 GHz and without higher resolution and higher sensitivity maps at longer wavelengths to fully characterize the SED. Similarly, we do not have ancillary data to discern if proto-stellar jets (point 3) are contributing to the non-thermal emission.
In the absence of lower frequency observations to rule out a turnover in the spectral index, higher resolution observations on scales of 1 -- 2 pc to further disentangle these crowded regions, or observations to probe the kinematics of these regions, we leave this question for future research. For this analysis, we use the 3 component SED and retain the classification from \citet{Cabanac:2005} for these knots.

Table \ref{tab:finalmodel} summarizes these results and gives the modeled, thermal flux density (S$_{ff,fit}$) at 33 GHz from the SED model. The uncertainties are propagated from the fit. Finally, we note that a more robust statistical analysis and SED model would ultimately bring clarity to these classifications; however, given the legitimate concerns about the 5, 15, and 33 GHz calibrations, we believe that such an analysis is not warranted at this time, and we leave it to future work when a multiwavelength observing campaign with better matched synthesized beams can be undertaken.

\begin{table}[htb!]
\centering
\caption{Summary of Best Fit SED Model for Radio Knots\label{tab:finalmodel}}
\begin{tabular}{P{1cm}P{5cm}P{4cm}P{2cm}P{2cm}}
\toprule
\multirow{2}{*}{Knot} &SED & \multirow{2}{*}{Classification} & S$_{ff,fit}$ & S$_{dust,fit}$ \\
& Components & & (mJy)  & (mJy) \\
\midrule
1a & thermal+dust & UD \HII region &  0.25 $\pm$ 0.02 & 0.15 $\pm$ 0.08  \\
 1b & thermal+dust & UD \HII region & 0.24 $\pm$ 0.02 & 0.16 $\pm$ 0.07 \\
 2 & thermal+dust & UD \HII region & 0.47 $\pm$ 0.04 & 0.37 $\pm$ 0.14\\
 4a & non-thermal+thermal+dust & \HII region + SNR & 0.52 $\pm$ 0.11 & 0.27 $\pm$ 0.18\\
 4b & non-thermal+thermal+dust& \HII region + SNR & 0.61 $\pm$ 0.11 & 0.15 $\pm$ 0.14 \\
 5 & thermal+dust & UD \HII region & 0.52 $\pm$ 0.04 & 0.13 $\pm$ 0.12 \\
\bottomrule
 \end{tabular}
\end{table}

\subsection{Production Rate of Ionizing Photons and Mass of the Clusters \label{sec:qly}}
If there is a purely thermal estimate of the flux density, we can determine the stellar mass of the embedded clusters hosted in the radio knots. We first estimate the production rate of Lyman continuum photons, Q$_{Lyc}$, with the specific thermal radio luminosity, \(L_{therm} = 4\pi D^2 \times S_{\nu} \), where D is the distance in cm and S$_{\nu}$ is the flux density in erg s$^{-1}$ cm$^{-2}$ Hz$^{-1}$. Q$_{Lyc}$ is then
\begin{equation}
	Q_{Lyc} \geq 6.3 \times 10^{52} \left(\frac{T_e}{10^4 \textrm{ K}}\right)^{-0.45}\left(\frac{\nu}{\textrm{GHz}}\right)^{0.1}\left(\frac{L_{therm}}{10^{27}\textrm{ ergs s$^{-1}$ Hz$^{-1}$}}\right) \textrm{ s}^{-1},
	\label{eq:qlyman}
\end{equation}
where T$_e$ is the electron temperature in Kelvin and $\nu$ is frequency in GHz \citep{Condon:1992,Johnson:2003}. We adopt 10$^4$ K as the typical temperature of an \HII region. Given the concerns about the 33 GHz data from the 15B-197 program, we utilize the thermal flux density modeled from the SED fit listed in Table \ref{tab:finalmodel} to estimate Q$_{Lyc}$.

We estimate the stellar mass using Starburst99 models \citep{Leitherer:1999,Vazquez:2005, Leitherer:2010,Leitherer:2014}. The 2014 update to the Starburst99\footnote{\url{https://www.stsci.edu/science/starburst99/docs/default.htm}} package includes synthesizing stellar populations with rotating stars, which are more luminous than non-rotating stars. Additionally, we can select the metallicity, the choice of initial mass function (IMF), and the mass cutoffs to estimate Q$_{Lyc}$.  In the Starburst99 package, we synthesized Q$_{Lyc}$ values for \ohz{= 0.014} and \ohz{= 0.008}. This is motivated by the results of \citet{Cresci:2017}, who mapped the metallicity of He 2-10 from their MUSE (Multi Unit Spectroscopic Explorer) observations. They report a range of metallicities (\ohm{} 8.25 -- 8.55) across He 2-10, which is in agreement with other estimates that averaged over the galaxy (e.g., \ohm{8.55 $\pm$ 0.02}; \ohz{$\sim$ 0.0098}; \citealt{Esteban:2014}). These two cases represent limits in Q$_{Lyc}$ for natal clusters in He 2-10. We used a Kroupa IMF \citep{Kroupa:2001} with two components (1.3 and 2.3) and with mass cutoffs of 0.1, 0.5, 100 \Msun{} as well as a Salpeter IMF \citep{Salpeter:1955} with mass cutoffs of 1 and 100 \Msun. For a 10$^6$ \Msun{} cluster that is less than 1 Myr old, the values of Q$_{Lyc}$ vary less than a factor of 2 for the choice of the metallicity, IMF, and rotation. We select a representative value of Q$_{Lyc}$ $\approx$ 5.3 $\times$10$^{52}$ photons s$^{-1}$ for a 10$^6$ \Msun{} cluster, with a standard deviation of 1.2 $\times$10$^{52}$ photons s$^{-1}$ which we treat as a systematic uncertainty in our analysis. In Table 7, columns 2 and 3 list Q$_{Lyc,fit}$ and the cluster mass, M$_{fit}$. The uncertainties are propagated through the equations. Finally, we note that these Q$_{Lyc}$ estimates, and thus the mass estimates, are lower limits because the ionizing photons from the UD \HII regions can be absorbed by dust.

The \palpha{} flux density also provides a lower limit on Q$_{Lyc}$, as it traces ionized gas but can suffer from extinction in extremely dense and massive dust clouds. The thermal flux density can be estimated from the line flux following Equation (3) of \citet{Condon:1992}
\begin{equation}
\frac{S_{T}}{\textrm{mJy}} \approx  \left(\frac{1}{0.28}\right) \left(\frac{\textrm{F(H}\beta)}{10^{-12} \textrm{ erg/s/cm$^2$}} \right) \left(\frac{T_e}{10^4 \textrm{ K}}\right)^{0.52}\left(\frac{\nu}{\textrm{GHz}}\right)^{-0.1},
\label{eq:lineflux}
\end{equation}
where F(H$\beta$) is the H$\beta$ line flux. From photometry, we estimate the \palpha{} flux density for each source. The flux density is multiplied by the FWHM of the filter (Table \ref{tab:alldataprop}) to calculate the \palpha{} line flux, F(\palpha). The ratio of the \palpha{} line flux to the H$\beta$ line flux at T$_e$ = 10$^4$ K is 
\begin{equation}
\frac{\textrm{F(Pa}\alpha)}{\textrm{F(H}\beta)}  = 0.317,
\label{eq:ratio}
\end{equation}
for line intensities from Table 4.4 of \citet{Osterbrock:1989}, assuming Case B and n$_e$ = 10$^6$ cm$^{-3}$. With Equations (\ref{eq:qlyman}) -- (\ref{eq:ratio}), we can estimate Q$_{Lyc}$ at 33 GHz from F(\palpha), assuming an electron temperature of 10$^4$ K and optically thin gas. Column 4 of Table \ref{tab:propertiescontour} lists the calculated values of Q$_{Lyc,\textrm{Pa}\alpha}$ from the \palpha{} flux densities in Table \ref{tab:fluxcontour}. For any embedded cluster or line of sight that is heavily extincted, the mass calculated from the thermal radio flux density should be higher than the mass from the \palpha{} flux density. Any discrepancy between the values inferred from radio and \palpha{} can be used to constrain either the extinction of \palpha{} and/or excess emission due to a non-thermal component in the radio flux. Q$_{Lyc,fit}$ is higher than Q$_{Lyc,\textrm{Pa}\alpha}$ for the radio knots, indicating that the \palpha{} map is affected by extinction towards these lines of sight.

As discussed in Section \ref{sec:photnicmos}, the extracted flux density is sensitive to the choice of aperture and treatment of the background subtraction. The region near Knots 4a and 4b has optically visible sources as well as bright features in the \palpha{} map that are not associated with the peaks in the radio emission (Figure \ref{fig:nicmos}). Particularly for Knot 4b, an aperture of equivalent size to the ones used to extract the radio flux densities will include the brightest feature in the \palpha{} map. However, this feature is not coincident with the peak in the 33 GHz map and is coincident with optical sources, as seen in the H$\alpha$ and other HST maps in Figure 3 of \citet{Cabanac:2005}. Inclusion of this bright feature returns a negative extinction value for Knot 4b, which is unphysical. Because the 33 GHz data were of dubious quality, we rely on the SED model to provide the thermal flux density value for our radio mass estimate. Given the 22 GHz and 113 GHz flux densities, we believe the modeled thermal flux density to be reasonable. However, it is still possible that the SED model and extracted flux densities for this source may be suspect. If the bright feature near Knot 4b is associated with the radio source, then an underestimated thermal flux might explain this discrepancy.

\begin{table}[htb!]
\centering
\begin{threeparttable}
\caption{Calculated Q$_{Lyc}$, Mass, and Extinction Estimates for Radio Knots\label{tab:propertiescontour}}
\begin{tabular}{P{0.8cm}P{2cm}P{2cm}P{2cm}P{1cm}}
\toprule

\multirow{3}{*}{Knot} & Q$_{Lyc,fit}\tnote{a}$ & M$_{fit}$ & Q$_{Lyc,\textrm{Pa}\alpha}$\tnote{b} & \multirow{3}{*}{A$_{V}$}\\ 
& $\times$10$^{51}$ & $\times$10$^{5}$ & $\times$10$^{51}$ & \\ 
 & (photons s$^{-1})$ & (\Msun) &  (photons s$^{-1}$) & \\ 
 \midrule
1a  & 2.12 $\pm$ 0.18 & 0.40 $\pm$ 0.10 & 0.34 $\pm$ 0.26  & 14  \\ 
1b & 2.10 $\pm$ 0.18 & 0.40 $\pm$ 0.10 & 1.17 $\pm$ 0.24  & 4.5 \\ 
2  & 4.05 $\pm$ 0.34 & 0.77 $\pm$ 0.19 & 2.67 $\pm$ 0.27 & 3.2 \\ 
4a & 4.51 $\pm$ 0.92 & 0.86 $\pm$ 0.26 & 3.41 $\pm$ 0.28 & 2.2\\ 
4b & 5.25 $\pm$ 0.95 & 1.00 $\pm$ 0.29 & 3.38 $\pm$ 0.27  &  3.4 \\ 
5  & 4.51 $\pm$ 0.35 & 0.86 $\pm$ 0.21 & 0.60 $\pm$ 0.23 &  16 \\ 
 
\bottomrule
\end{tabular}
\begin{tablenotes}
\item[a] Calculated using the modeled thermal flux density at 33 GHz (S$_{ff,fit}$) in Table \ref{tab:finalmodel}.
\item[b] Calculated using the \palpha{} flux density in Table \ref{tab:fluxcontour}.
\end{tablenotes}
\end{threeparttable}
\end{table}

\subsection{Extinction\label{sec:extinction}}
The \palpha{} map may be affected by extinction due to extremely dense and massive dust clouds. We can estimate the extinction, A$_V$, by comparing the radio and \palpha{} results. \citet{Indebetouw:2005} present an analysis of the wavelength dependence of interstellar extinction between 1.25 -- 8.0 $\mu$m and describe the behavior of \(A_{\lambda}/A_K\) in this wavelength range, where A$_K$ is the infrared $K$ band (2.2 $\mu$m) extinction. From Equation (4) of \citet{Indebetouw:2005}, we expect A$_{\textrm{Pa}\alpha}$/A$_K$ $\sim$1.24  and for R$_V$ = 3.1, A$_V$/A$_K$ $\sim$8.8 \citep{Cardelli:1989}. If we assume the radio Q$_{Lyc,fit}$ value is purely thermal and any decrement in Q$_{Lyc}$ between the radio and \palpha{} values is due to extinction, then 
\begin{equation}
\textrm{A}_V \approx -17.7 \log\left(\frac{\textrm{Q}_{Lyc, Pa\alpha}}{\textrm{Q}_{Lyc, fit}}\right).
\end{equation}
The last column of Table \ref{tab:propertiescontour} gives the estimates of the extinction values towards the radio knots, which is between 2.2 -- 16. \citet{Cabanac:2005} find an extinction of A$_V$ = 10.5 from the ratio of Br$\gamma$/Br10 using the line fluxes from \citet{Vanzi:1997}. The observations of \citet{Vanzi:1997} were long slit observations with the 23 m Bok telescope of Steward Observatory with an aperture of 2.4\arcsec{} $\times$ 15.6\arcsec. Our estimates of the extinction are more localized but the range is consistent with the extinction reported by \citet{Cabanac:2005}. Given that both the \palpha{} and radio flux densities are lower limits, the extinction values we derive are also lower limits and the actual values could be higher. We discuss the variation in the extinction between the radio knots in Section \ref{sec:dis}.

\subsection{Submillimeter Emission, Dust Mass, and Total Mass\label{sec:dustmass}}

The 340 GHz flux density can provide a probe of the dust mass associated with the radio knots. Star forming galaxies can have dust temperatures between 25 -- 40 K but active galaxies can have higher T$_d$ \citep{Scoville:2014}. \citet{Gorski:2017} report a gas temperature near forming SSCs in NGC 253 of 130 K. In the absence of short wavelength observations of similar spatial resolution to the radio data to determine the dust tempertature, we adopt a range of dust temperatures of T$_d$ = (25, 40, 100) K, which are similar to the values discussed in \citet{Vacca:2002} for cocoons of dust enveloping SSCs in He 2-10. Though, \citet{Vacca:2002} note that a single blackbody is a not a good representation of the dust SED between 10.8 -- 100 $\mu$m.  From the SED models, we can estimate the flux density due to the dust emission, S$_{dust,fit}$, that is not contaminated by free-free or non-thermal emission. Column 5 of Table \ref{tab:finalmodel} lists these values. Comparing the modeled dust emission, S$_{dust,fit}$ to the measured flux density at 340 GHz, we estimate that 48 -- 68$\%$ of the 340 GHz flux density is due to free-free emission. This agrees with other estimates from \citet{Hirashita:2013}, who extrapolated the \citet{Johnson:2003} 5 GHz flux density to 340 GHz and estimated from their 340 GHz observations that 45 -- 64$\%$ of the 340 GHz emission was due to free-free emission. With S$_{dust,fit}$, we calculate the specific luminosity, L$_{\nu}$, and then the dust mass, M$_d$, is

\begin{equation}
\textrm{M}_d = \frac{\textrm{L}_{\nu}}{4\pi\,\kappa_{\nu}\,\textrm{B}_{\nu}(\textrm{T}_d)},
\end{equation}
where the value of B$_{\nu}$(T$_{d}$) is in erg s$^{-1}$ cm$^{-2}$ Hz$^{-1}$ from the Planck function at 340 GHz. The absorption coefficient of dust is  \[\kappa_{\nu} = \kappa_o \left(\frac{\nu}{\nu_o}\right)^{\beta}, \] where $\nu$  = $\nu_o$ in this case and \(\beta = \alpha_{dust} - 2\). We adopt $\kappa_o$ = 1.3 cm$^2$ g$^{-1}$ at $\nu_o$ = 340 GHz \citep{Whitcomb:1981,Hildebrand:1983} but note that the uncertainty can be as large as a factor of 2. Table \ref{tab:dustmass} gives the calculated dust mass for T$_d$ = (25, 40, 100) K. Adopting a dust-to-gas ratio of 1:100 \citep{Wilson:2008} and assuming M$_{fit}$ is the stellar contribution to the total mass, the knots have total masses between 0.5 -- 1.1 $\times$ 10$^5$ \Msun{} for T$_d$ = 100 K. For colder dust, the total mass estimate will increase slightly.

\begin{table}[htb!]
\centering
\begin{threeparttable}
\caption{Dust Mass \label{tab:dustmass}}
\begin{tabular}{P{1cm}P{2cm}P{2cm}P{2cm}}
\toprule

\multirow{2}{*}{Knot}  & \multicolumn{3}{c}{M$_d$ (10$^3$ \Msun)} \\\cmidrule{2-4}
& T$_d$ = 25 K & T$_d$ = 40 K & T$_d$ = 100 K\\
\midrule
1a & 0.7 & 0.4 & 0.1 \\ 
1b & 0.7 & 0.4 & 0.1 \\
2 & 1.6 & 0.9 & 0.3 \\
4a & 1.2 & 0.7 & 0.2 \\
4b  & 0.7 & 0.4 & 0.1 \\
5 & 0.6 & 0.3 & 0.1 \\
\bottomrule
\end{tabular}
\end{threeparttable}
\end{table}

\subsection{Star Formation Efficiency \label{sec:SFE}}
A broader interest in SSCs is as a local analog to globular clusters. SSCs must remain bound in order to evolve into a globular cluster, and the star formation efficiency (SFE) of the cluster is one measure of the cluster's ability to remain bound. If most of the cluster's mass is in the form of gas, then when the gas is expelled from the cluster through stellar feedback mechanisms, the cluster may no longer remain bound. The threshold to remain bound, as traced by the SFE, has varied in the literature over the years. One conservative estimate is SFE $>$ 90$\%$ \citep{Hills:1980}, but \citet{Baumgardt:2007} and \citet{Pelupessy:2012} suggest that it could be as low as 5 -- 10$\%$ if the timescale over which the gas is removed is long. Alternatively, \citet{Geyer:2001} suggest SFE $>$ 50$\%$ in order to form bound clusters. \citet{Matthews:2018} provide a review of the SFE in this context and in the merging environment of the Antennae galaxies, which also host SSCs. In the Antennae galaxies, they found few clusters with SFE $>$ 20$\%$ but note that their analysis was not sensitive to embedded clusters.

Following the treatment of the SFE from \citet{Matthews:2018}, the instantaneous mass ratio is 
\begin{equation}
\textrm{IMR(t)} = \left(1+\frac{M_{*}}{M_{gas}}\right)^{-1},
\end{equation}
where M$_{*}$ is the stellar mass and M$_{gas}$ is the gas mass. The IMR is a function of time over the evolution of the cluster. When the cluster is still forming stars, the IMR $<$ 1, and when star formation ceases but the natal molecular cloud is not yet disrupted, the IMR approaches unity and is then equal to the SFE. With S$_{ff,fit}$ as an estimate for M$_*$ and S$_{dust,fit}$ with a 1:100 dust-to-gas ratio for M$_{gas}$, we can estimate the IMR for the radio knots. For T$_d$ = 100 K, the IMR is between $\sim$71 -- 88$\%$ for the radio knots; for colder dust (25 K), the IMR is $\sim$32 -- 60$\%$. For a larger dust-to-gas ratio, the IMR will decrease; for example, a dust-to-gas ratio of 1:300 yields IMR $\sim$45 -- 72$\%$ for T$_d$ = 100 K and IMR $\sim$ 14 -- 33$\%$ for T$_d$ = 25 K. For warm dust and a 1:100 dust-to-gas ratio, high IMR suggests that the natal SSCs in He 2-10 could potentially remain bound even after the gas is dispersed through stellar feedback processes. If the dust is cold or the dust-to-gas ratio is larger however, the clusters may not remain bound.

\subsection{Discussion\label{sec:dis}}
\begin{figure}[htb!]
\centering

\gridline{\fig{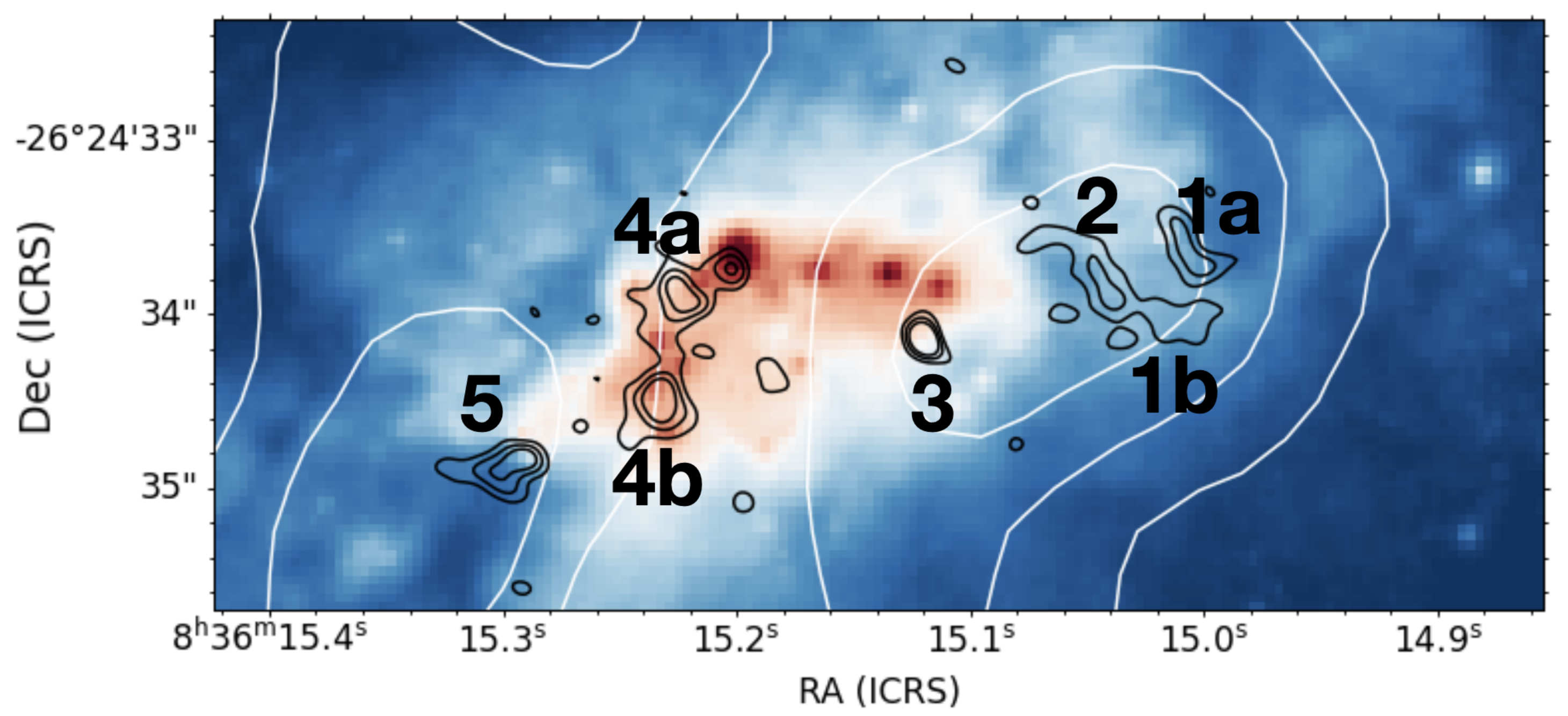}{0.49\textwidth}{(a) F814W, $^{12}$CO(2-1), $\&$ 22 GHz}}

\gridline{\fig{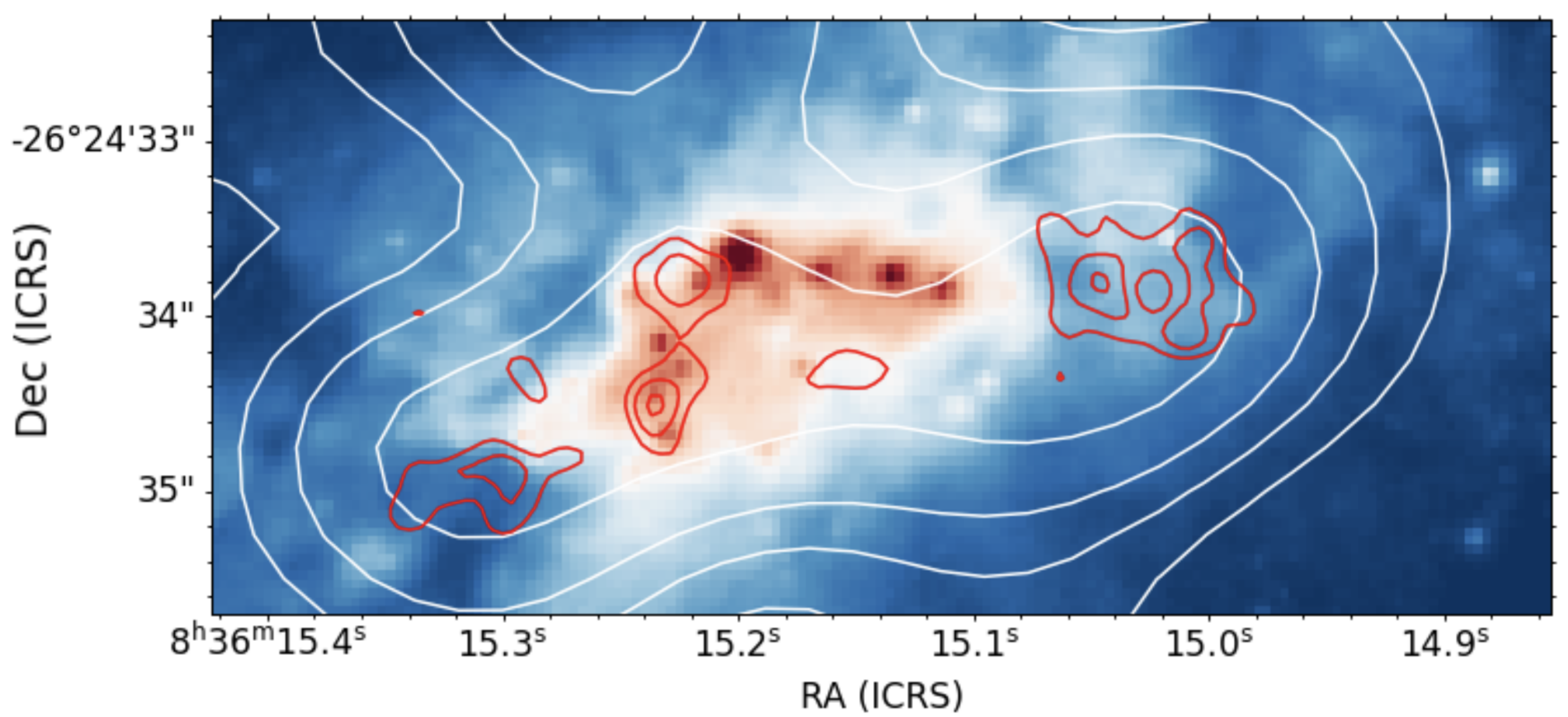}{0.49\textwidth}{(b) F814W, HCO$^+$(1-0), $\&$ 340 GHz}}

\gridline{\fig{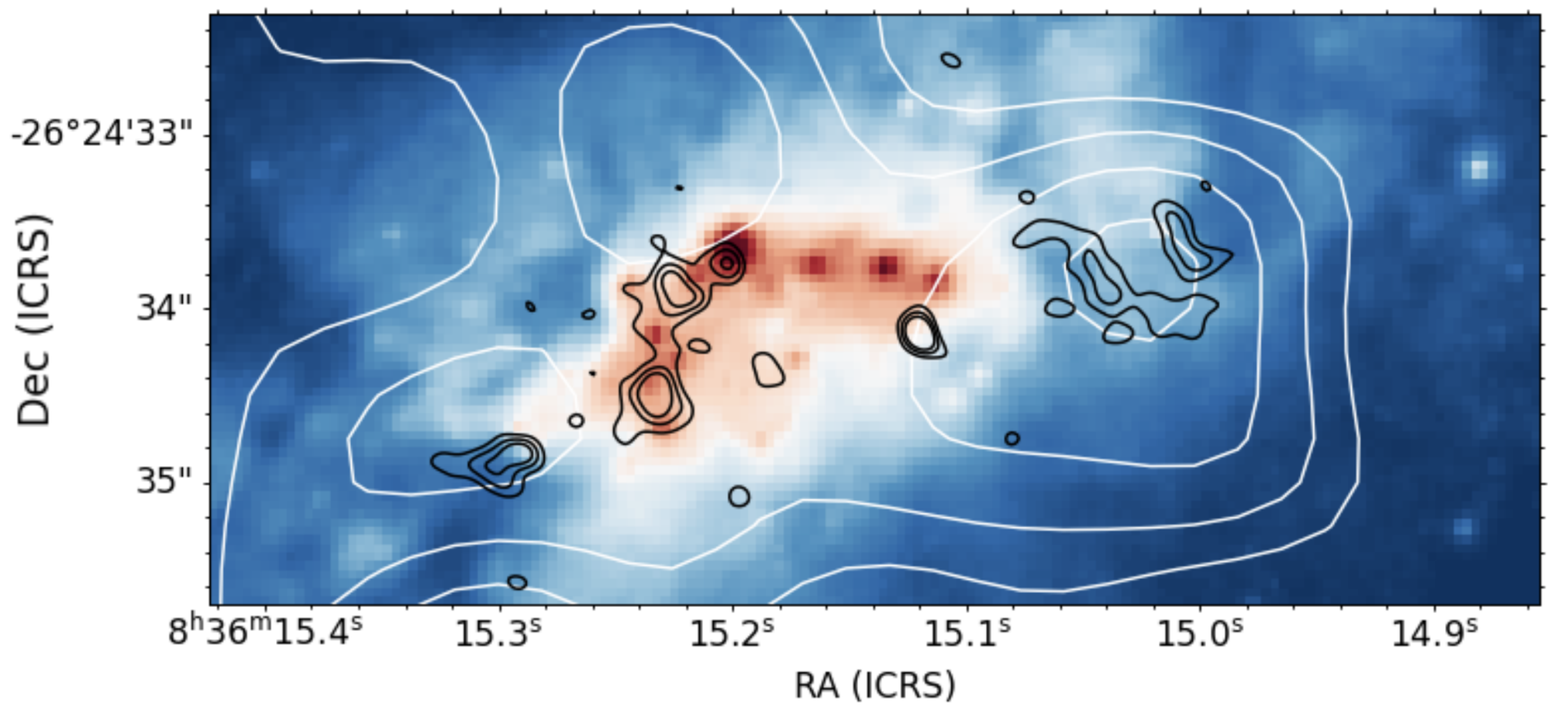}{0.49\textwidth}{(c) F814W, HCN(1-0), $\&$ 22 GHz}}

\gridline{\fig{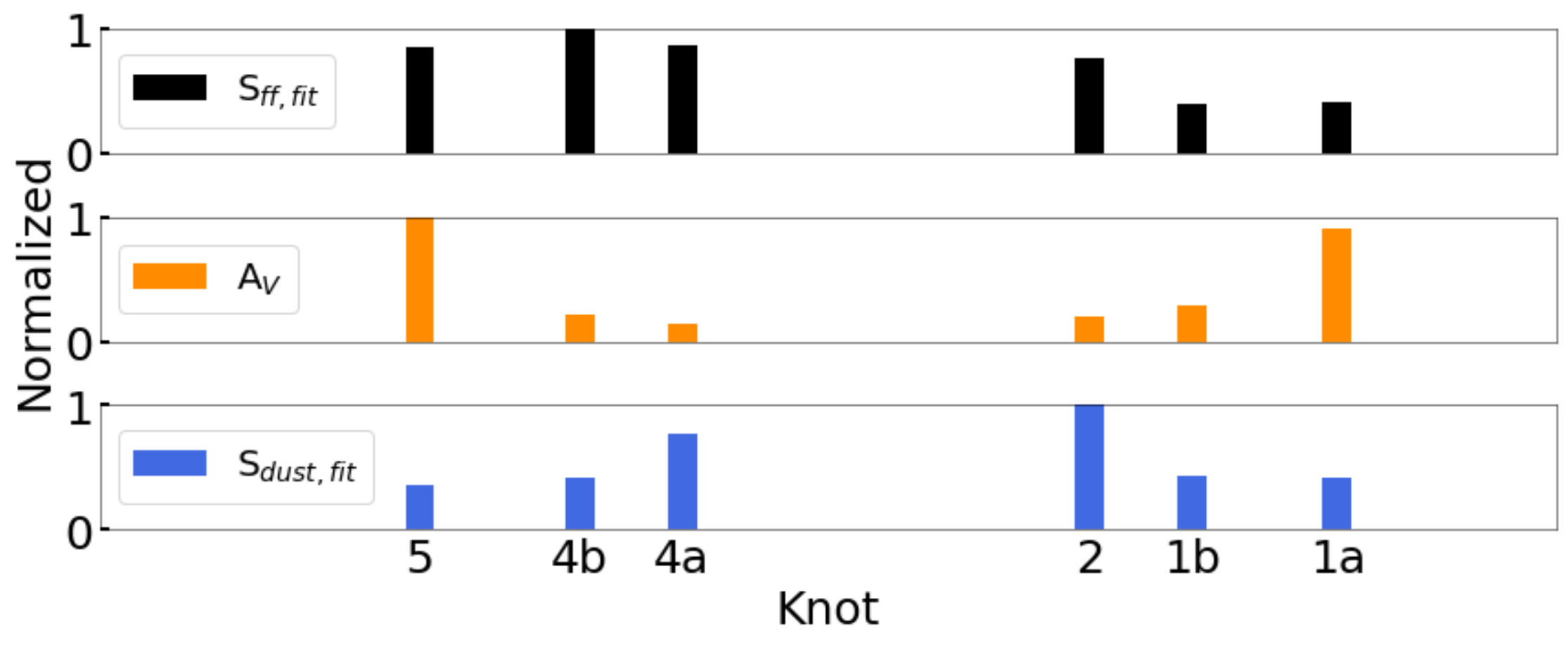}{0.49\textwidth}{(d)}}
\caption{In the first three panels, the raster map is the HST/WFPC2 F814W filter, showing the optical morphology of He 2-10. In (a), the black contours are the unconvolved VLA 22 GHz continuum data at 3, 5, and 7$\sigma$. The white contours are SMA $^{12}$CO(2-1) moment 0 (1.8\arcsec{} $\times$ 1.2\arcsec{}) at 3, 5, and 7$\sigma$ ($\sigma$ = 0.02 Jy beam$^{-1}$ km s$^{-1}$). In (b), the red contours are the unconvolved ALMA 340 GHz continuum data at 3, 5, and 7$\sigma$, and the white contours are ALMA HCO$^+$(1-0) moment 0 (1.7\arcsec{} $\times$ 1.6\arcsec) at 3, 5, 7, and 10$\sigma$ ($\sigma$ = 1.9 Jy beam$^{-1}$ km s$^{-1}$). In (c), the black contours are VLA 22 GHz, and the white contours are ALMA HCN(1-0) moment 0 (1.7\arcsec{} $\times$ 1.6\arcsec{}) at 2, 3, 4, and 5$\sigma$ ($\sigma$ = 0.02 Jy beam$^{-1}$ km s$^{-1}$). The three panels of (d) are bar plots of the normalized thermal radio brightness (S$_{ff,fit}$; black), extinction (A$_V$; orange), and the submillimeter brightness (S$_{dust,fit}$; blue) for each knot. Each parameter set is normalized by the maximum value.}
\label{fig:normpar}
\end{figure}

From the analysis of the SEDs for the radio knots, we have presented estimates of the production rate of ionizing photons, the extinction towards the radio knots from the \palpha{} and thermal radio emission, and the emission associated with dust, as traced by the 340 GHz data. The bottom three panels of Figure \ref{fig:normpar} display these parameters with respect to their spatial distribution across He 2-10; the top panels show a HST/WFPC2 F814W raster map with 22\footnote{We use the unconvolved 22 GHz map to trace the thermal emission instead of the 33 GHz map in Figure \ref{fig:normpar} because of the calibration concerns in the 33 GHz data. The morphology in the 22 and 33 GHz maps is similar, as seen by comparing the contours in Figure \ref{fig:nicmos}.} and 340 GHz continuum in black and red contours, respectively, and the white contours show (\ref{fig:normpar}a) SMA $^{12}$CO(2-1), (\ref{fig:normpar}b) ALMA HCO$^+$(1-0), and (\ref{fig:normpar}c) ALMA HCN(1-0) velocity-integrated intensity (moment 0) maps from \citet{Johnson:2018}. With these results, we discuss here general trends as traced by the radio knots in He 2-10.

The western and eastern regions of He 2-10, as traced by Knots 1a and 5 have higher A$_V$ values but are not submm bright. In the absence of observations at shorter wavelengths to determine the dust temperature, the simple assumption is that T$_d$ is the same for each knot; however, the geometry of the dust with respect to the cluster impacts the observed submm brightness and A$_V$. Given the low submm emission but higher A$_V$ values, the dust could be cold and far from these clusters, perhaps in form of a foreground dust screen; though, the dust could also be much hotter and peaking at shorter wavelengths. From their HNC(1-0) and HCN(1-0) observations of the natal SSCs, \citet{Johnson:2018} report that the molecular material near Knots 1a, 1b, and 2 are the warmest in the region while the region near Knot 5 is cooler. If the dust and molecular material is cospatial, then the observations and analysis presented here are not sensitive to the peak of the dust emission near Knots 1a, 1b, and 2. The closely spaced lines of sight to Knots 1a, 1b, and 2 and the variation in A$_V$ between them suggests that the dust distribution may be patchy in this region as well. The knots are also associated with peaks in the CO(2-1) and HCN(1-0) emission. CO is a ``generic'' molecular gas tracer; it is easily photodissociated by young massive stars, requiring only 11.09 eV to do so. HCN(1-0), however, is a cold, dense gas tracer with a critical density of $n_{\textrm{crit}}$ = 1.7 $\times$ 10$^5$ cm$^{-3}$ at 50 K for optically thin gas \citep{Shirley:2015}. These clusters appear to be consistent with being embedded and young, as previously reported by \citet{Johnson:2003} and \citet{Cabanac:2005}, and Knot 5 may be the youngest and most embedded of the group.

Knots 4a and 4b are bright in the submm and have the lowest A$_V$ values. The dust is unlikely to be in a foreground screen but instead close to the cluster and warm. The submm emission closely follows the morphology of the thermal emission, indicating that the submm emission is not a spatially extended feature of this region. The evolved SSCs in the vicinity of these knots complicates the interpretation of the region however. We included a non-thermal component in the SED models for these knots, but this could be due to contamination from the evolved SSCs in the vicinity. The contaminating non-thermal emission would lower the estimated thermal flux, which in turn decreases A$_V$. Because of the lack of optical counterparts coincident with the radio emission, these knots appear to be consistent with being embedded clusters. However, they are not associated with peaks in the CO(2-1) and HCN(1-0) maps, suggesting that the knots are not as embedded, or as young, as Knots 1a, 1b, and 5. Alternatively, these knots could be of a similar age if the more evolved SSCs in the vicinity were almost entirely responsible for destroying the natal material in this region.

The central region of He 2-10, as traced by Knots 4a and 4b and the optical clusters, appears to be in a distinct evolutionary state compared to the most eastern and western sides, where the latter two may host younger, more embedded clusters. Previous studies have noted an age gradient across He 2-10 (e.g., \citealt{Johnson:2000}, \citealt{Cabanac:2005}, \citealt{Chandar:2003}). Knot 4a\footnote{The studies of \citet{Johnson:2000}, \citet{Johnson:2003}, \citet{Cabanac:2005}, and \citet{Chandar:2003} do not use the nomenclature of Knot 4a, and in the latter two, they do not necessary assess properties of the radio knots directly, focusing on the nearby optical and IR sources instead. We specifically are referring to radio sources, which lack optical counterparts, when we use the ``knot'' terminology.  In the crowded regions of Knot 4a and 4b, the resolutions of these older studies and the apertures the authors selected likely include both radio knots and optical sources. Age estimates that rely on optical and IR tracers, like those of \citet{Cabanac:2005} and \citet{Chandar:2003}, may not be accurate for the radio knots and comparison of the optical clusters and radio sources needs to be done carefully, as we have tried to do here.} may have an age $>$ 6 Myr  \citep{Johnson:2003, Cabanac:2005}, though \citet{Chandar:2003} suggest that the optical cluster near Knot 4a may be $<$ 6 Myr because of a P Cygni profile associated with that cluster, which puts a upper limit on the age. Knots 1 (as identified in the lower resolution maps), 2, and 5 are younger by comparison, with ages $\leq$ 5 Myr \citep{Cabanac:2005}.

\begin{figure}[htb!]
\centering
      \includegraphics[width=0.7\textwidth]{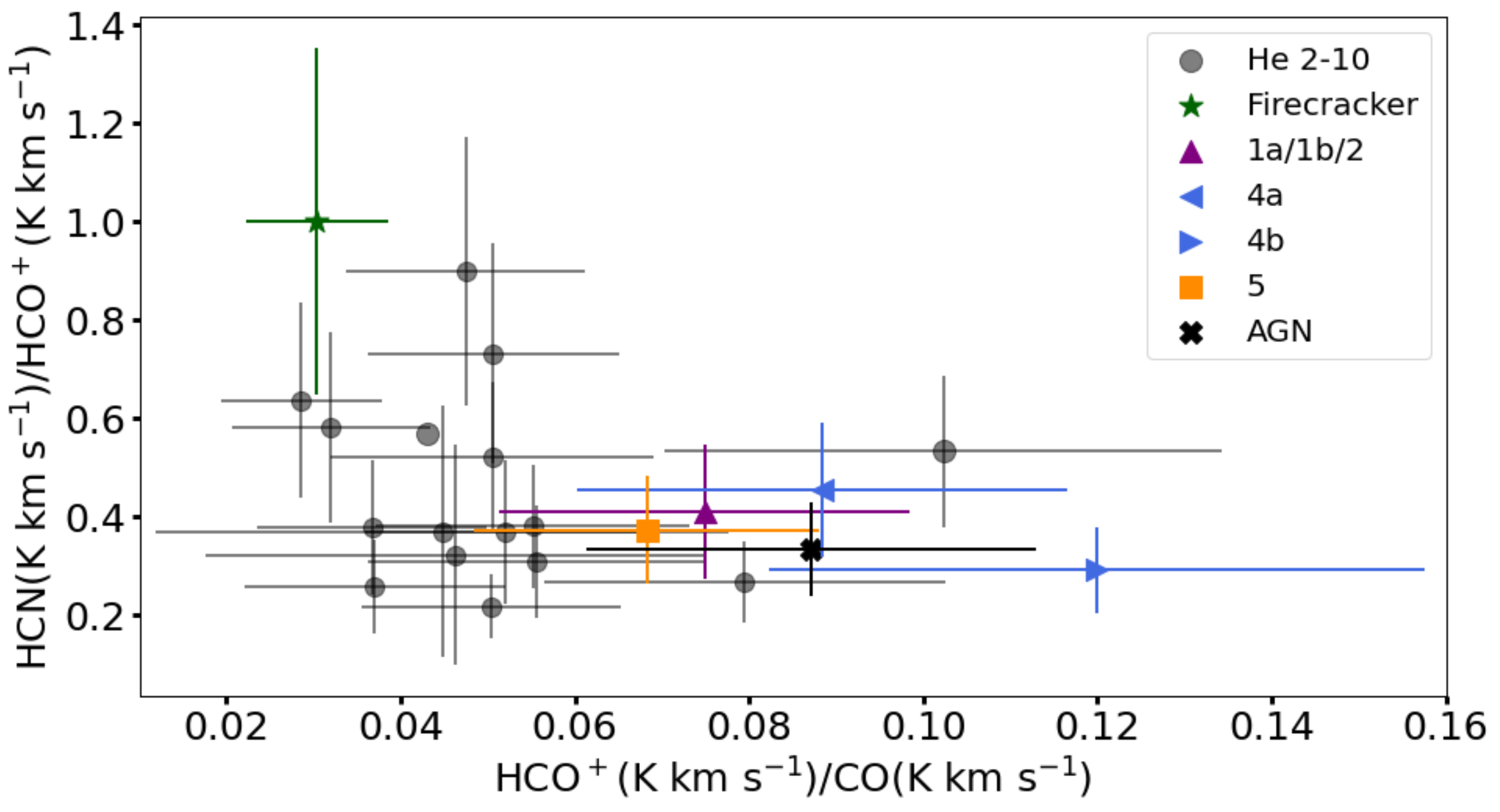}
\caption{Ratio of HCN/HCO$^+$ and HCO$^+$/CO for molecular clouds in He 2-10 (\citealt{Johnson:2018}; gray circles) and the Firecracker in the Antennae galaxies (\citealt{Finn:2019}; green star). The sources from \citet{Johnson:2018} that correspond to the radio knots are represented with unique symbols as given in the legend. The HCO$^+$/CO ratio for the Firecracker is a lower limit.}
\label{fig:molratios}
\end{figure}

\citet{Johnson:2018} report HCN(1-0), HCO$^+$(1-0), and CO(2-1) line strengths for nineteen 0.8\arcsec{} apertures, chosen to probe peaks in their ALMA and SMA maps of He 2-10. The HCO$^+$(1-0) line traces warmer, moderately dense gas ($n_{\textrm{crit}}$ = 2.9 $\times$ 10$^4$ cm$^{-3}$; \citealt{Shirley:2015}) and is commonly associated with \pdr{s.} HCO$^+$ does not seem to vary across the radio knots, unlike the CO and HCN maps. He 2-10 appears to have moderately dense, warm gas present in the central region, while the colder, denser gas traced by HCN(1-0) is more patchy and peaks preferentially on the edges of the central region. From a principle component analysis of these lines and others for clouds in He 2-10, \citet{Johnson:2018} suggest that the ratios of HCN/HCO$^+$ and HCO$^+$/CO are correlated with the evolutionary state of the cluster.  In Figure \ref{fig:molratios}, we show an adaptation of Figure 13 from \citet{Johnson:2018} and Figure 13 of \citet{Finn:2019} to show this trend. The emerging clusters are at the bottom right (higher HCO$^+$/CO, low HCN/HCO$^+$) and as the HCO$^+$/CO ratio decreases and HCN/HCO$^+$ increases, the clusters are thought to be more embedded and younger. At the upper left of the plot are the precursor molecular clouds that have the potential to form SSCs but have not yet begun doing so. An example of such a cloud is the Firecracker in the Antennae galaxies \citep{Finn:2019}.

In Figure \ref{fig:molratios}, we highlight the apertures from \citet{Johnson:2018} that are associated with the radio knots. Knots 4a and 4b appear in the bottom right and Knot 5 is to the left of them, which is consistent with our interpretation of the degree to which these knots are embedded. With the resolution of the HCN(1-0), HCO$^+$(1-0), and CO(2-1) maps ($\sim$1.7\arcsec, $\sim$74 pc), Knots 1a, 1b, and 2 are in a single beam and we cannot differentiate physical conditions between the knots in the ALMA maps. From the available continuum data, Knot 1a is similar to Knot 5, so we expect that in higher resolution data, this knot would be closer to Knot 5 in this figure. The curiosity is Knot 2, however, in that the A$_V$ is low and it is submm bright like Knots 4a and 4b; however, it is associated with peaks in the HCN(1-0) and CO(2-1) maps, suggesting that it is embedded like Knots 1a, 1b, and 5. Higher resolution observations of the dense gas tracers will be able to disentangle the characteristics of these knots further. Generally, our analysis of the non-thermal, thermal, and dust emission presented here supports the empirical correlation between the molecular ratios of HCN/HCO$^+$ and HCO$^+$/CO and the evolutionary state of the cluster as suggested by \citet{Johnson:2018}.

\subsection{Concluding Remarks}

With higher resolution data and additional wavelength coverage, we are beginning to move away from the early simple models of the formation of SSCs. Complexities are emerging, which may include amorphous morphologies, non-instantaneous star formation, etc. As we probe finer spatial scales, we may find smaller \HII regions, corresponding to a cluster formation scenario that is hierarchical and fragmented instead of one that undergoes monolithic collapse. This may present, for example, as a bound cluster composed of UC \HII regions instead of a single diffuse \HII region encompassing the cluster. As future telescopes such as the Next Generation VLA (ngVLA) and the James Space Webb Telescope (JWST) are able to probe smaller spatial scales and provide complementary multiwavelength observations, studies will offer new insights and constraints on star formation theory.

\section{Summary\label{sec:sum}}

We present observations of He 2-10 from 1.87 $\mu$m to 6 cm to constrain the properties of 7 radio knots in the most complete multiwavelength survey on parsec scales to date of He 2-10. Our conclusions are as follows.

\begin{itemize}
\item We present maps at 5, 8, 15, 22, 33, 113, 250, and 340 GHz and 1.87 $\mu$m of He 2-10. We discuss photometry results for the radio knots, which were initially identified in \citet{Kobulnicky:1999} and \citet{Johnson:2003}. With the higher resolution of these maps compared to previous studies, we extract individual clusters from aggregate complexes as identified at lower resolution. In total, we identify seven knots: Knots 1a, 1b, 2, 3, 4a, 4b, and 5. Knot 3 is known to host the AGN \citep{Reines:2011, Reines:2012, Reines:2016}.

\item We construct spectral energy distributions from radio to submillimeter wavelengths and model the SED to disentangle the non-thermal, thermal, and dust components. The key figures and table showing these results are Figures \ref{fig:contour1} and \ref{fig:contour2} and Table \ref{tab:finalmodel}. Knots 1a, 1b, 2, and 5 are consistent with being UD \HII regions, as initially concluded by \citet{Johnson:2003} and \citet{Cabanac:2005}. The SEDs for Knots 4a and 4b appear to have a non-negligible non-thermal component. These knots may be a mixture of normal \HII regions and SNRs as suggested by \citet{Cabanac:2005} or the evolved SSCs in the vicinity may be contributing the contaminating non-thermal emission. These knots are not associated with optical counterparts, so the clusters do not appear to have completely emerged yet from their natal material.

\item From the SED model, we calculate the rate of ionizing photons from the thermal flux density, which can provide an estimate for the mass of the cluster, and we find stellar masses between 0.4 -- 1.0 $\times$ 10$^5$ \Msun{} for the natal SSCs hosted in the radio knots. The range of masses for the natal SSCs is consistent with estimates of stellar masses for adolescent SSCs in He 2-10 from \citet{Johnson:2000}. From the HST/NICMOS F187N observations, the \palpha{} flux densities for the radio knots provide a lower limit on the ionizing fluxes as well. \palpha{} can suffer from extinction in extremely dense and massive dust clouds, so we estimate the extinction along the line of sight by comparing the radio and \palpha{} estimates. We find A$_V$ values between 2.2 -- 16 for the radio knots. Table \ref{tab:propertiescontour} lists these values. 

\item In Section \ref{sec:dustmass}, we estimate the dust content associated with the knots from the SED models to be 0.1 -- 0.3 $\times$ 10$^3$ \Msun{} at 100K. If the dust temperature is cooler, the dust mass estimates will increase. Assuming a  dust-to-gas ratio of 1:100 and the stellar mass estimated from the thermal flux density, we find total masses between 0.5 -- 1.1 $\times$ 10$^5$ \Msun{}. In Section \ref{sec:SFE}, we estimate the instantaneous mass ratio (IMR), which is related to the star formation efficiency, and find a range of 71 -- 88$\%$ for the radio knots. For 100 K dust and a 1:100 dust-to-gas ratio, high IMR suggests that the natal SSCs in He 2-10 could potentially remain bound even after the gas is dispersed through stellar feedback processes, perhaps allowing these clusters to evolve into objects similar to globular clusters. If the dust is cold or the dust-to-gas ratio is larger however, the clusters may not remain bound. 

\item In Section \ref{sec:dis}, we discuss characteristics of the radio knots based on the results of this study and previous ones to construct a holistic picture of the radio knots in He 2-10. Knots 4a and 4b appear to be still marginally embedded, as they do not have optical counterparts, and the dust near these knots is likely compact and warm. Knots 1a and 5 are characterized by high A$_V$ values but relatively little emission at 340 GHz, suggesting that they are embedded and cold dust may exist in a patchy, foreground screen. Generally, the central region of He 2-10, as traced by Knots 4a and 4b and the optical SSCs, appears to be more evolved than the eastern and western regions. We compare our interpretation to ALMA maps of dense gas tracers such as CO, HCN, and HCO$^+$ in Figures \ref{fig:normpar} and \ref{fig:molratios}. Our multiwavelength analysis supports the empirical correlation between these dense gas tracers and the cluster evolution scenario proposed by \citet{Johnson:2018}.
\end{itemize}

\acknowledgements

This paper makes use of the following ALMA data: ADS/JAO. ALMA$\#$2012.1.00413.S., 2015.1.01569.S, 2016.1.00027.S, and 2016.1.00419.S. ALMA is a partnership of ESO (representing its member states), NSF (USA) and NINS (Japan), together with NRC (Canada), NSC and ASIAA (Taiwan), and KASI (Republic of Korea), in cooperation with the Republic of Chile. The Joint ALMA Observatory is operated by ESO, AUI/NRAO and NAOJ. The National Radio Astronomy Observatory is a facility of the National Science Foundation operated under cooperative agreement by Associated Universities, Inc. AER gratefully acknowledges support for this work provided by NASA through EPSCoR grant number 80NSSC20M0231. This research made use of Photutils, an Astropy package for detection and photometry of astronomical sources \citep{Bradley:2019}. The majority of this paper was written during the 2020 lockdown in the workspace of home, so we thank those around us who patiently supported and encouraged us. Finally, we thank the referee of this paper for a helpful and collegial review

\facility{VLA, ALMA}

\newpage
\bibliographystyle{aasjournal}
\bibliography{FRbib,LBVbib,compbib,FollowupBib,Biblio,Antennaebib}

\end{document}

%% file: defs.tex
\newcommand{\vlatext}{The Karl G. Jansky Very large Array is an instrument of the National Radio Astronomy Observatory (NRAO). The NRAO is a facility of the National Science Foundation, operated under cooperative agreement with Associated Universities, Inc.}
\newcommand{\ghz}{$\;$GHz}

\newcommand{\palpha}{Pa$\alpha$}
\newcommand{\ohm}{12 + log(O/H) = }
\newcommand{\ohz}{Z }

\newcommand{\ALMA}{Atacama Large Millimeter/submillimeter Array (ALMA)}

\newcommand{\pdr}{photodissociation region}

\newcommand{\I}{\textit{I}}
\newcommand{\Q}{\textit{Q}}
\newcommand{\U}{\textit{U}}

\newcommand{\HII}{H\,{\sc ii} } 
\newcommand{\ddeg}{\ensuremath{^{\circ}}} 

\newcommand{\Msun}{\ensuremath{\textrm{M}_{\odot} }} 

\newcommand{\kms}{km s$^{-1}$}